\documentclass[aps,prev,twocolumn,preprintnumbers,floatfix,nofootinbib]{revtex4-2}

\pdfoutput=1
\usepackage[T1]{fontenc}
\usepackage[utf8]{inputenc}
\usepackage{graphicx}
\usepackage{xcolor}
\usepackage{amsmath,amssymb,amsfonts}
\usepackage{booktabs}
\usepackage{multirow}
\usepackage{array}
\usepackage{slashed}
\usepackage{bm}
\usepackage{hyperref}
\usepackage{tikz}
\usepackage{tikz-feynman}
\usepackage{adjustbox}

\tikzfeynmanset{compat=1.1.0}

\hypersetup{
	colorlinks=true,
	linkcolor=blue,
	citecolor=blue,
	urlcolor=blue
}

\providecommand{\openone}{\leavevmode\hbox{\small1\kern-3.8pt\normalsize1}}

\newcommand{\RE}{\mathrm{Re}}

\begin{document}
\preprint{IFJPAN-IV-2026-7}

\title{Probing Heavy Neutral Higgs Bosons via Single Vector-Like Bottom Quark Production at the HL-LHC}

\author{R. Benbrik$^{1}$}
\email{r.benbrik@uca.ac.ma}

\author{M. Berrouj$^{1}$}
\email{mbark.berrouj@ced.uca.ma}

\author{M. Boukidi$^{2}$}
\email{mohammed.boukidi@ifj.edu.pl}

\author{M. Ech-chaouy$^{1}$}
\email{m.echchaouy.ced@uca.ac.ma}

\author{K. Kahime$^{3}$}
\email{Kahimek@gmail.com}

\author{K. Salime$^{1}$}
\email{k.salime.ced@uca.ac.ma}

\affiliation{\vspace{0.25cm}$^1$Polydisciplinary Faculty, Laboratory of Physics, Energy, Environment, and Applications, Cadi Ayyad University, Sidi Bouzid, B.P. 4162, Safi, Morocco.\\
	$^2$Institute of Nuclear Physics, Polish Academy of Sciences, ul. Radzikowskiego 152, Cracow, 31-342, Poland.\\
	$^3$Laboratoire Interdisciplinaire de Recherche en Environnement, Management, Energie et Tourisme (LIREMET), ESTE, Cadi Ayyad University, B.P. 383, Essaouira, Morocco.}
	\begin{abstract}
		We investigate the discovery prospects of a singly produced vector-like bottom quark in the Type-II Two-Higgs-Doublet Model extended by an $SU(2)_L$ vector-like $(T,B)$ doublet. We focus on the non-standard decay chain $B \to \phi b$, followed by $\phi \to t\bar{t}$, where $\phi = H$ or $A$, leading to a final state with one charged lepton, missing transverse energy, and multiple $b$-jets. We perform a full simulation of both signal and Standard Model backgrounds at $\sqrt{s}=14$ TeV. We show that the exotic channels $B \to \phi b$ can dominate over the conventional decay modes, reaching branching ratios of order $50\%$ for both neutral scalars in the alignment limit. A conventional cut-based analysis provides a $5\sigma$ discovery significance only at sufficiently high integrated luminosity. By contrast, an XGBoost-based multivariate analysis substantially improves the signal-background discrimination and extends the discovery reach up to $m_B \simeq 1.3$ TeV with $600~\mathrm{fb}^{-1}$ and up to $m_B \simeq 1.6$ TeV with $3~\mathrm{ab}^{-1}$, even in the presence of systematic uncertainties as large as $15\%$. 
	\end{abstract}
	
	\maketitle
	
	\section{Introduction}
	
	The Standard Model (SM) has achieved remarkable success in describing particle interactions up to the electroweak scale. Nevertheless, several open issues, including the hierarchy problem and the possible non-minimal structure of the scalar sector, strongly motivate extensions beyond the SM \cite{DeSimone:2012fs}. Among the most compelling and widely studied possibilities are models with extra dimensions \cite{Chang:1999nh,Gherghetta:2000qt,Contino:2003ve}, Little Higgs constructions \cite{Arkani-Hamed:2002iiv,Schmaltz:2002wx,Chang:2003vs,Han:2003wu}, composite Higgs scenarios \cite{Agashe:2004rs,Bellazzini:2014yua,Contino:2006qr,Lodone:2008yy,Matsedonskyi:2012ym}, and grand unified theories \cite{Hewett:1988xc}. A common prediction of many such frameworks is the existence of vector-like quarks (VLQs). These are color-triplet spin-$1/2$ fermions whose left- and right-handed components transform identically under the SM gauge group $SU(3)_C \times SU(2)_L \times U(1)_Y$. Unlike the chiral quarks of the SM, VLQs can therefore acquire gauge-invariant masses independently of electroweak symmetry breaking. Depending on their electroweak representation, they may appear as singlets [$(T)$, $(B)$], doublets [$(T,B)$, $(X,T)$, $(Y,B)$], or triplets [$(X,T,B)$, $(T,B,Y)$], and can couple to SM quarks through Yukawa interactions involving the scalar sector \cite{delAguila:2000aa}. In many realistic constructions, these couplings are dominated by the third generation, leading to the conventional decay channels $B \to Wt$, $B \to Zb$, and $B \to hb$ for a down-type vector-like bottom quark~\cite{Nutter:2012an, Atre:2011ae,Okada:2012gy, Backovic:2014uma, Yang:2025zbp, Han:2025itd, Yang:2024aav, Cao:2022mif, Liu:2024hvp, Zhang:2024nto, Han:2025itd}.
	
	A particularly economical extension of the scalar sector is provided by the Two-Higgs-Doublet Model (2HDM) \cite{Gunion:1992hs,Branco:2011iw}, in which the SM Higgs sector is enlarged by a second $SU(2)_L$ scalar doublet. After electroweak symmetry breaking, the physical spectrum contains two CP-even neutral states, $h$ and $H$, one CP-odd neutral state, $A$, and a pair of charged Higgs bosons, $H^\pm$. Such a scalar structure arises naturally in a number of well-motivated BSM settings, including supersymmetric frameworks~\cite{Moretti:2019ulc}, and gives rise to a rich collider phenomenology. When VLQs are embedded into the 2HDM, the presence of additional scalar states opens non-standard decay channels such as $Q \to Hq$, $Q \to Aq$, and $Q \to H^\pm q$ \cite{Gopalakrishna:2015wwa,Benbrik:2019zdp,Benbrik:2024hsf,Arhrib:2024nbj,Benbrik:2023xlo,Benbrik:2025nfw,Benbrik:2022kpo,Benbrik:2024bxt,Arhrib:2024mbq,Arhrib:2024dou,Arhrib:2024tzm,Arhrib:2016rlj,Abouabid:2023mbu,Angelescu:2015uiz,Ghosh:2023xhs,Dermisek:2019vkc, Benbrik:2026zjv, Dermisek:2020gbr, Dermisek:2021zjd, Banerjee:2016wls}. Once kinematically open, these exotic modes can acquire sizable branching fractions and may even dominate over the conventional SM channels. As a consequence, the decay pattern of VLQs can differ significantly from the simplified assumptions commonly adopted in experimental searches, and the resulting collider bounds may be substantially weakened. This strongly motivates dedicated searches specifically targeting Higgs-mediated VLQ decays \cite{Benbrik:2025kvz,Bhardwaj:2022nko,Roy:2020fqf, BENBRIK2026117436}.
	
	The LHC has performed extensive searches for VLQs at center-of-mass energies of 7, 8, and 13 TeV, probing both pair and single production for several electroweak representations. In particular, searches for singly produced vector-like bottom quarks have placed stringent constraints on the relevant effective couplings, excluding values up to $k_B \sim s_R^d \lesssim 0.3$ \cite{Buchkremer:2013bha,Matsedonskyi:2014mna,ATLAS:2023ixh,CMS:2024bni,Benbrik:2024fku}. However, these limits are generally derived under the restrictive assumption that the heavy quark decays only into SM final states. More recently, the CMS Collaboration has investigated singly produced vector-like top quarks decaying into non-standard final states such as $t\phi$, where $\phi$ denotes a neutral scalar that can be identified with $H$ or $A$ in the context of the 2HDM+VLQ framework \cite{CMS:2025zwi}. Although no statistically significant excess has been observed, these analyses illustrate the growing experimental interest in heavy-quark decays mediated by an extended scalar sector.
	
	Motivated by this, we study the single production of a vector-like bottom quark followed by the exotic decay $B \to \phi b$, with $\phi = H$ or $A$, and the subsequent decay $\phi \to t\bar{t}$. We focus on the semileptonic final state in which one top quark decays leptonically, yielding a signature with one charged lepton, missing transverse energy, and multiple $b$-jets. This channel simultaneously probes the heavy-quark and heavy-Higgs sectors and is particularly well suited for multivariate analyses. We show that, in realistic regions of parameter space consistent with current constraints, the branching ratio $\mathrm{BR}(B \to \phi b)$ can reach $\mathcal{O}(50\%)$, making this channel one of the most promising discovery modes. We perform a full collider study at $\sqrt{s}=14$ TeV and demonstrate that an analysis based on XGBoost significantly outperforms a conventional cut-based strategy, substantially extending the discovery reach at the HL-LHC.
	
	The remainder of the paper is organized as follows. In Sec.~\ref{sec:Framework}, we summarize the theoretical framework. In Sec.~\ref{sec:Constraints}, we discuss the theoretical and experimental constraints. In Sec.~\ref{sec:cutbased}, we present the cut-based analysis, while Sec.~\ref{sec:ML} contains the machine-learning setup and the main numerical results. We conclude in Sec.~\ref{sec:Conclusions}.
	
	\section{Theoretical framework}
	\label{sec:Framework}
	
	In the 2HDM with a softly broken $Z_2$ symmetry, the scalar sector contains two complex $SU(2)_L$ doublets, $\Phi_1$ and $\Phi_2$. Assuming CP conservation, the most general renormalizable and gauge-invariant scalar potential can be written as \cite{Branco:2011iw,Gunion:1989we}
	\begin{eqnarray}
		V\left( \Phi_1, \Phi_2 \right)  &=& m_{11}^2 \Phi_1^\dag \Phi_1 + m_{22}^2 \Phi_2^\dag \Phi_2 - m_{12}^2 \left( \Phi_1^\dag \Phi_2 + \Phi_2^\dag \Phi_1 \right) \nonumber \\
		&+& \frac{\lambda_1}{2} \left( \Phi_1^\dag \Phi_1 \right)^2 + \frac{\lambda_2}{2} \left( \Phi_2^\dag \Phi_2 \right)^2 \nonumber \\
		&+& \lambda_3 \left( \Phi_1^\dag \Phi_1 \right) \left( \Phi_2^\dag \Phi_2 \right) + \lambda_4 \left( \Phi_1^\dag \Phi_2 \right) \left( \Phi_2^\dag \Phi_1 \right) \nonumber\\
		&+& \frac{\lambda_5}{2} \left[ \left( \Phi_1^\dag \Phi_2 \right)^2 + \left( \Phi_2^\dag \Phi_1 \right)^2 \right],
		\label{thdmV}
	\end{eqnarray}
	where all parameters are real.
	
	Passing to the Higgs basis, only one linear combination of the two doublets acquires a nonzero vacuum expectation value,
	\begin{eqnarray}
		H_1 = \left( \begin{array}{c}
			G^+ \\
			\frac{v + \varphi^0_1 + i G^0}{\sqrt{2}} \\
		\end{array} \right), \quad
		H_2 = \left( \begin{array}{c}
			H^+ \\
			\frac{\varphi^0_2 + i A}{\sqrt{2}} \\
		\end{array} \right),
	\end{eqnarray}
	with $v = \sqrt{v_1^2 + v_2^2} \simeq 246$ GeV. Here, $G^0$ and $G^\pm$ denote the Goldstone bosons and $H^\pm$ is the charged Higgs boson. The CP-even neutral fields $\varphi^0_1$ and $\varphi^0_2$ mix into the mass eigenstates $h$ and $H$ according to
	\begin{eqnarray}
		\left( \begin{array}{c}
			h \\
			H \\
		\end{array} \right)
		=
		\left( \begin{array}{cc}
			\sin(\beta - \alpha) & \cos(\beta - \alpha) \\
			\cos(\beta - \alpha) & -\sin(\beta - \alpha) \\
		\end{array} \right)
		\left( \begin{array}{c}
			\varphi_1^0 \\
			\varphi_2^0 \\
		\end{array} \right),
	\end{eqnarray}
	where $\tan\beta = v_2/v_1$ and $\alpha$ diagonalizes the CP-even scalar mass matrix. In the alignment limit, $\sin(\beta-\alpha)\to 1$, the lighter CP-even state $h$ reproduces the properties of the SM Higgs boson.
	
	The possible VLQ representations under $SU(3)_C \times SU(2)_L \times U(1)_Y$ are
	\begin{align}
		& T^0_{L,R}, \quad B^0_{L,R} && \text{(singlets)} \,, \notag \\
		& (X,T^0)_{L,R}, \quad (T^0,B^0)_{L,R}, \quad (Y,B^0)_{L,R} && \text{(doublets)} \,, \notag \\
		& (X,T^0,B^0)_{L,R}, \quad (T^0,B^0,Y)_{L,R} && \text{(triplets)} \,.
	\end{align}
	The superscript $0$ denotes weak-eigenstate fields. In the present study, we focus on a vector-like bottom quark belonging to the $(T,B)$ doublet. The state $B$ has electric charge $Q=-1/3$ and mixes with the SM bottom quark after electroweak symmetry breaking.
	
	The presence of $B^0_{L,R}$ enlarges the down-quark sector to four mass eigenstates, $d$, $s$, $b$, and $B$. Owing to the stringent flavor and electroweak constraints, the mixing is assumed to occur predominantly with the third generation, as motivated, for example, by the LEP measurements of $R_b$ \cite{Aguilar-Saavedra:2002phh}. The mixing between $b^0$ and $B^0$ is parametrized as
	\begin{eqnarray}
		\left( \begin{array}{c}
			b_{L,R} \\
			B_{L,R} \\
		\end{array} \right)
		=
		\left( \begin{array}{cc}
			\cos\theta_{L,R}^d & -\sin\theta_{L,R}^d e^{i \phi_d} \\
			\sin\theta_{L,R}^d e^{-i \phi_d} & \cos\theta_{L,R}^d \\
		\end{array} \right)
		\left( \begin{array}{c}
			b^0_{L,R} \\
			B^0_{L,R} \\
		\end{array} \right),
		\label{ec:mixd}
	\end{eqnarray}
	where $\theta_{L,R}^d$ are the left- and right-handed mixing angles and $\phi_d$ is a CP-violating phase, which we neglect throughout this analysis.
	
	In the Higgs basis, the relevant Yukawa structure is
	\begin{equation}
		-\mathcal{L}_Y \supset y^u \bar{Q}^0_L \tilde{H}_2 u^0_R + y^d \bar{Q}^0_L H_1 d^0_R + M_u^0 \bar{u}_L^0 u_R^0 + M_d^0 \bar{d}_L^0 d_R^0 + \text{h.c.},
	\end{equation}
	with $u_R^0=(u_R,c_R,t_R,T_R)$ and $d_R^0=(d_R,s_R,b_R,B_R)$. The mass matrix in the vector-like bottom sector takes the form
	\begin{eqnarray}
		\mathcal{L}_\text{mass} = - \left( \begin{array}{cc}
			\bar{b}_L^0 & \bar{B}_L^0
		\end{array} \right)
		\left( \begin{array}{cc}
			y_{33}^d \frac{v}{\sqrt{2}} & y_{34}^d \frac{v}{\sqrt{2}} \\
			y_{43}^d \frac{v}{\sqrt{2}} & M^0
		\end{array} \right)
		\left( \begin{array}{c}
			b_R^0 \\
			B_R^0
		\end{array} \right) + \text{h.c.},
		\label{ec:Lmass}
	\end{eqnarray}
	where $M^0$ is the vector-like bare mass and $y_{ij}^d$ are Yukawa couplings. Diagonalization is performed through the bi-unitary transformation
	\begin{equation}
		U_L^d \mathcal{M}^d (U_R^d)^\dagger = \mathcal{M}^d_\text{diag}.
		\label{ec:diag}
	\end{equation}
	
	The mixing angles satisfy
	\begin{eqnarray}
		\tan 2 \theta_L^d &=& \frac{\sqrt{2} |y_{34}^d| v M^0}{(M^0)^2 - \frac{1}{2} v^2(|y_{33}^d|^2 + |y_{34}^d|^2)} \quad \text{(singlets, triplets)}, \notag \\
		\tan 2 \theta_R^d &=& \frac{\sqrt{2} |y_{43}^d| v M^0}{(M^0)^2 - \frac{1}{2} v^2(|y_{33}^d|^2 + |y_{43}^d|^2)} \quad \text{(doublets)},
		\label{ec:angle1}
	\end{eqnarray}
	together with
	\begin{eqnarray}
		\tan \theta_R^q &=& \frac{m_q}{m_Q} \tan \theta_L^q \quad \text{(singlets, triplets)}, \notag \\
		\tan \theta_L^q &=& \frac{m_q}{m_Q} \tan \theta_R^q \quad \text{(doublets)}.
		\label{ec:rel-angle1}
	\end{eqnarray}
	
	In the alignment limit, the partial width for the decay of the vector-like bottom quark into a neutral scalar state is given by
	\begin{align}
		\Gamma(B \to \phi\, b) &= \frac{g^2}{128\pi}\,\frac{m_B}{M_W^2}\,\lambda^{1/2}(m_B,m_b,M_\phi)
		\nonumber\\
		&\times \left[
		(|Y^L_{\phi bB}|^2 + |Y^R_{\phi bB}|^2)(1+r_b^2-r_\phi^2)
		\pm 4r_b\,\RE\!\left(Y^L_{\phi bB}Y^{R*}_{\phi bB}\right)
		\right],
		\label{eq:GammaB}
	\end{align}
	where the plus sign corresponds to the CP-even state $H$ and the minus sign to the CP-odd state $A$, with $\phi=H,A$. The coefficients $Y^{L,R}_{\phi bB}$ denote the corresponding chiral couplings. Neglecting the suppressed right-handed contribution, one obtains in the alignment limit
	$
	Y^L_{\phi bB} = \tan\beta\, s_R^u s_R^d,
	$
	for $\sin(\alpha-\beta)=1$\cite{Arhrib:2024dou}.
	
	For $m_B \gtrsim 1~\mathrm{TeV}$, bottom-mass effects are negligible, i.e. $r_b \simeq 0$. Moreover, if $m_A \simeq m_H$, the decay widths become approximately equal,
	$
	\Gamma(B \to Ab) \approx \Gamma(B \to Hb) \approx \Gamma(B \to \phi b).
	$
	
	\subsection{Experimental and theoretical constraints}
	\label{sec:Constraints}
	
	We impose a set of theoretical and experimental constraints in order to identify viable regions of parameter space consistent with perturbativity, electroweak precision data, Higgs measurements, and direct collider bounds.
	
	\subsubsection*{Theoretical constraints}
	
	\begin{itemize}
		\item \textbf{Unitarity:} The $S$-wave amplitudes for scalar-scalar, scalar-gauge, and gauge-gauge scattering are required to satisfy perturbative unitarity at high energies \cite{Kanemura:1993hm}.
		
		\item \textbf{Perturbativity:} All quartic couplings in the scalar potential are required to obey $|\lambda_i| < 8\pi$ for $i=1,\dots,5$ \cite{Branco:2011iw}.
		
		\item \textbf{Vacuum stability:} The scalar potential must remain bounded from below in all field directions. This implies \cite{Deshpande:1977rw,Barroso:2013awa}
		\begin{align}
			&\lambda_1 > 0,\quad \lambda_2 > 0,\quad \lambda_3 > -\sqrt{\lambda_1 \lambda_2}, \notag \\
			&\lambda_3 + \lambda_4 - |\lambda_5| > -\sqrt{\lambda_1 \lambda_2}.
		\end{align}
		
		\item \textbf{Electroweak precision observables:} The oblique parameters $S$ and $T$ \cite{Grimus:2007if} are required to satisfy the 95\% confidence-level bounds from the global electroweak fit, assuming $U=0$ \cite{ParticleDataGroup:2020ssz},
		\begin{align}
			S = 0.05 \pm 0.08,\quad T = 0.09 \pm 0.07,\quad \rho_{ST}=0.92.
		\end{align}
		In the presence of VLQs, the total contribution is evaluated through $\chi^2(S_{\text{2HDM}}+S_{\text{VLQ}},\,T_{\text{2HDM}}+T_{\text{VLQ}})$. The VLQ corrections are computed using the analytic results of Ref.~\cite{Arhrib:2024tzm}. All constraints are implemented with a modified version of \texttt{2HDMC-1.8.0} \cite{Eriksson:2009ws}, including the VLQ effects as discussed in Refs.~\cite{Benbrik:2022kpo,Abouabid:2023mbu}.
	\end{itemize}
	
	\subsubsection*{Experimental constraints}
	
	\begin{itemize}
		\item \textbf{Searches for additional Higgs bosons:} The heavy neutral states $H$ and $A$, as well as the charged state $H^\pm$, are constrained using \texttt{HiggsBounds-6} \cite{Bechtle:2008jh,Bechtle:2011sb,Bechtle:2013wla,Bechtle:2015pma} within the \texttt{HiggsTools} framework \cite{Bahl:2022igd}, ensuring consistency with exclusion limits from LEP, Tevatron, and the LHC.
		
		\item \textbf{SM-like Higgs measurements:} Compatibility with the observed 125 GeV Higgs boson is tested with \texttt{HiggsSignals-3} \cite{Bechtle:2020uwn,Bechtle:2020pkv}, also through \texttt{HiggsTools}, requiring $\Delta\chi^2 \le 6.18$ at 95\% confidence level over 159 signal-strength measurements.
		
		\item \textbf{$b \to s\gamma$:} In the Type-II 2HDM, the radiative decay $b \to s\gamma$ yields the well-known lower bound $m_{H^\pm} \gtrsim 580$ GeV. In the presence of VLQs, this limit can be relaxed through loop-induced cancellations. For instance, in the $(TB)$ doublet scenario, viable configurations with $m_{H^\pm}\sim 360$ GeV may arise depending on the mixing pattern \cite{Benbrik:2022kpo}. In this work, we conservatively impose $m_{H^\pm} \ge 600$ GeV.
		
		\item \textbf{LHC constraints on VLQs:} Bounds from direct LHC searches are imposed by requiring $\sigma_{\text{theo}}/\sigma_{\text{obs}} < 1$ for the vector-like bottom quark, following the procedure detailed in Ref.~\cite{Benbrik:2024fku}.
	\end{itemize}
	
	\subsection{Production and decay of vector-like bottom quarks}
	
\begin{figure}[t]
		\centering
		\begin{minipage}{0.48\columnwidth}
			\centering
			\begin{tikzpicture}[scale=0.55]
				\begin{feynman}
					\vertex (g) at (-1, -3) {\(g\)};
					\vertex (q) at (-1, 3) {\(q\)};
					\vertex (v_g) at (1, -1.5);
					\vertex (v_q) at (1, 0.5);
					\vertex (v_T1) at (2.2, -0.5);
					\vertex (v_T2) at (3.2, -0.5);
					\vertex (q_prime) at (5, 3) {\(q'\)};
					\vertex (b_bar_1) at (5, -3) {\(\bar{b}\)};
					\vertex (b_2) at (5, -1.5) {\(b\)};
					\vertex (v_H) at (5, 0.6);
					
					\diagram* {
						(g) -- [gluon] (v_g),
						(q) -- [fermion] (v_q),
						(v_g) -- [anti fermion] (b_bar_1),
						(v_g) -- [fermion, edge label'=\(b\)] (v_T1),
						(v_q) -- [fermion] (q_prime),
						(v_q) -- [photon, edge label=\(Z\)] (v_T1),
						(v_T1) -- [fermion, green!60!black, ultra thick, edge label=\(\mathbf{B}\)] (v_T2),
						(v_T2) -- [fermion] (b_2),
						(v_T2) -- [scalar, red, edge label'=\(\phi\)] (v_H),
					};
				\end{feynman}
			\end{tikzpicture}
		\end{minipage}
		\hfill
		\begin{minipage}{0.48\columnwidth}
			\centering
			\begin{tikzpicture}[scale=0.55]
				\begin{feynman}
					\vertex (q) at (-3, 2.5) {\(q\)};
					\vertex (g) at (-3,-2.5) {\(g\)};
					\vertex (vZ)  at (-1.0, 0.5);
					\vertex (vBB) at (-1,-1.5);
					\vertex (vBbar) at (1, -0.5);
					\vertex (vB) at (1, -2.5);
					\vertex (qprime) at (4, 2.5) {\(q'\)};
					\vertex (bZ) at (4, 1.0) {\(b\)};
					\vertex (bA) at (3,-3.5) {\(b\)};
					\vertex (A)  at (3, -1.5);
					
					\diagram*{
						(q) -- [fermion] (vZ),
						(vZ) -- [fermion] (qprime),
						(vZ) -- [photon, edge label=\(Z\)] (vBbar),
						(g) -- [gluon] (vBB),
						(vBB) -- [anti fermion, green!60!black, ultra thick, edge label=\(\mathbf{\bar B}\)] (vBbar),
						(vBB) -- [fermion, green!60!black, ultra thick, edge label'=\(\mathbf{B}\)] (vB),
						(vBbar) -- [fermion] (bZ),
						(vB) -- [fermion] (bA),
						(vB) -- [scalar, red, dashed, edge label=\(\phi\)] (A),
					};
				\end{feynman}
			\end{tikzpicture}
		\end{minipage}
		\caption{Representative Feynman diagrams for single production of a vector-like bottom quark followed by the decay into a heavy neutral scalar $\phi$.}
		\label{fig:Bprod}
	\end{figure}
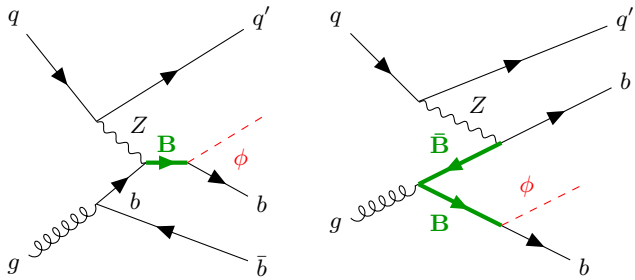

	The single production of a vector-like bottom quark at the LHC proceeds dominantly through electroweak interactions, primarily via $Zb$ fusion, as illustrated in Fig.~\ref{fig:Bprod}. The produced heavy quark subsequently decays into a bottom quark and a heavy neutral scalar, $\phi = H, A$.
	
\begin{figure*}[t]
		\centering
		\includegraphics[width=\textwidth]{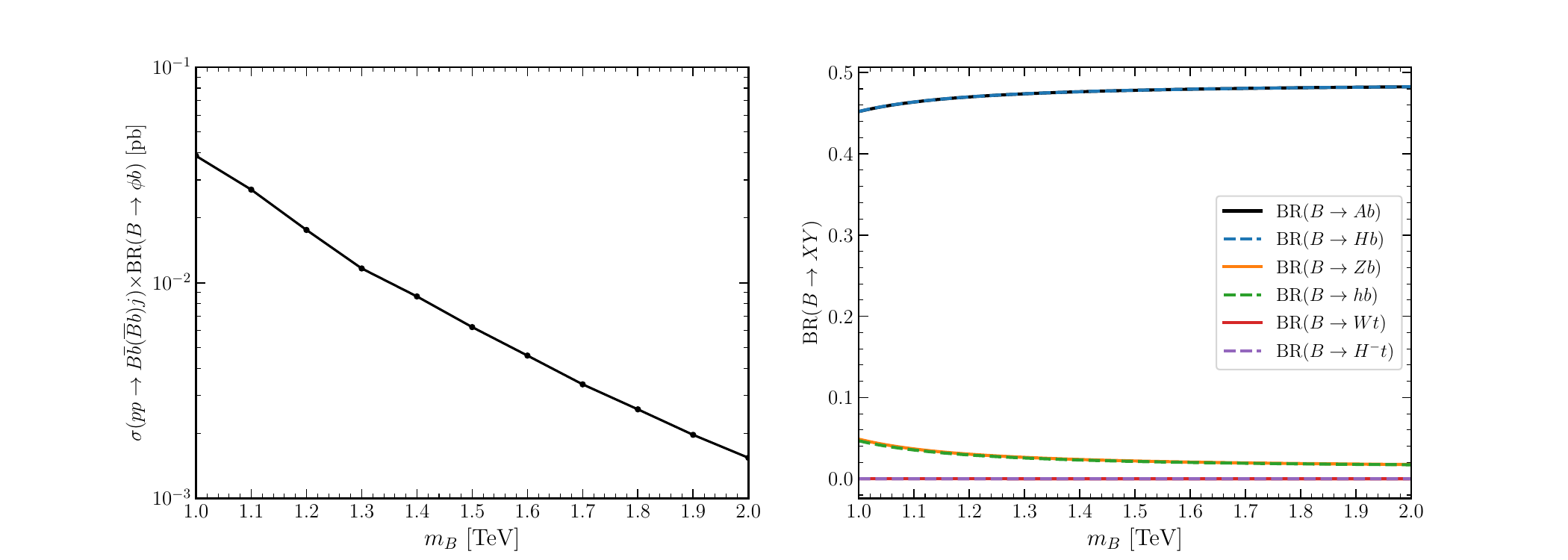}
		\caption{Left: leading-order $\sigma\left(pp \to \overline{B}b(B\overline{b})j\right)\times\mathrm{BR}\left(B(\overline{B}) \to \phi b(\overline{b})\right)$ with $\phi=H,A$ as a function of $m_B$. Right: corresponding branching ratios as functions of $m_B$, for $m_H=m_A=m_{H^\pm}=700$ GeV, $\tan\beta=6$, $s_R^u=0.01$, and $s_R^d=0.2$.}
		\label{fig:xs_br}
	\end{figure*}

	In the left panel of Fig.~\ref{fig:xs_br}, we show
	$\sigma(pp \to \overline{B}b\,(B\overline{b})j)\times \mathrm{BR}(B(\overline{B}) \to \phi\, b(\overline{b}))$
	as a function of the vector-like bottom mass $m_B$. The cross section decreases from about $0.04~\mathrm{pb}$ at $m_B=1~\mathrm{TeV}$ to about $1.5\times10^{-3}~\mathrm{pb}$ at $m_B=2~\mathrm{TeV}$, reflecting the expected phase-space suppression at larger masses.
	
	The right panel displays the corresponding branching ratios as functions of $m_B$ for the representative benchmark point
	$m_H = m_A = m_{H^\pm} = 700~\mathrm{GeV}$, $\tan\beta = 6$, $s_R^u = 0.01$, and $s_R^d = 0.2$.
	In this setup, the exotic decays into heavy neutral Higgs bosons dominate, with
	$\mathrm{BR}(B \to H b) \simeq \mathrm{BR}(B \to A b) \simeq 48\%$
	over almost the full mass range. By contrast, the conventional channels $B \to Z b$ and $B \to h b$ remain suppressed at the level of $\sim 1\%$, while the charged channels $B \to W^- t$ and $B \to H^- t$ are absent for the chosen parameter point.
	
	This pattern can be directly understood from the underlying coupling structure. The $\phi Bb$ interaction scales as $\propto \tan\beta\, s_R^d c_R^d$, which enhances the neutral non-standard channels. By contrast, the $BZb$ and $Bhb$ couplings do not receive a $\tan\beta$ enhancement and scale only as $\propto s_R^d c_R^d$, leading to a strong suppression. The charged couplings $BW^- t$ and $BH^- t$ are further suppressed by the hierarchy $s_R^d \gg s_R^u$, which effectively closes these decay modes in the region of interest~\cite{BENBRIK2026117436}.
	
	\section{Cut-based analysis}
	\label{sec:cutbased}
	
	We select benchmark points satisfying all theoretical and experimental constraints. In particular, we choose $s_R^d \gg s_R^u$ in order to enhance the signal rate, since ${\cal BR}(B \to \phi b) \propto s_R^d c_R^d$.
	
	The selected benchmark points listed in Table~\ref{tab:BPs} satisfy
	$
	\left|{\cal BR}(B\to Ab) - {\cal BR}(B\to Hb)\right| \lesssim 5\%,
	\qquad
	\left|{\cal BR}(H\to t\bar{t}) - {\cal BR}(A\to t\bar{t})\right| \lesssim 5\%.
	$
	Hence, the signal cross sections associated with the two heavy neutral Higgs bosons are nearly identical. A dedicated analysis of one of the two states therefore leads to essentially the same sensitivity as the other. For this reason, we present the study in terms of a generic heavy neutral scalar $\phi$.
	
	\begin{table}[t]
		\centering
		\begin{adjustbox}{width=0.7\columnwidth}
			\begin{tabular}{l|c|c|c}
				\hline\hline
				Parameters & BP$_1$ & BP$_2$ & BP$_3$ \\
				\hline
				$m_h$   & 125.09 & 125.09 & 125.09 \\
				$m_H$ & 730.6 & 492.11 & 559.75 \\
				$m_A$  & 745.28 & 576.06 & 585.77 \\
				$m_{H^\pm}$ & 792.21 & 613.01 & 748.51 \\
				$t_\beta$ & 3.8 & 1.58 & 1.59 \\
				$s_{\beta-\alpha}$ & 1 & 1 & 1 \\
				$m_T$ & 1218.11 & 1299.81 & 1699.61 \\
				$m_B$ & 1222.96 & 1342.37 & 1755.26 \\
				$s^u_L$ & 0.0014 & 0.0013 & 0.0010 \\
				$s^d_L$ & 0.0006 & 0.0008 & 0.0006 \\
				$s^u_R$ & 0.01 & 0.01 & 0.01 \\
				$s^d_R$ & 0.2 & 0.2 & 0.2 \\
				$\Gamma_B$ & 106.56 & 84.06 & 52.65 \\
				\hline
				\multicolumn{4}{c}{$\mathcal{BR}$ in \%} \\
				\hline
				${\cal BR}(B\to Ab)$  & 41.17 & 29.94 & 32.84 \\
				${\cal BR}(A\to t\bar{t})$  & 84.79 & 99.29 & 99.84 \\
				${\cal BR}(B\to Hb)$  & 43.64 & 33.71 & 33.57 \\
				${\cal BR}(H\to t\bar{t})$  & 80.1 & 98.87 & 99.14 \\
				\hline\hline
			\end{tabular}
		\end{adjustbox}
		\caption{Benchmark points for the 2HDM+$TB$ scenario. Masses and widths are given in GeV.}
		\label{tab:BPs}
	\end{table}
	
	The signal process under consideration is
	\begin{equation}
		pp \to B\bar{b}j \to \phi b\bar{b}j \to t\bar{t}b\bar{b}j 
		\to W^{+} b\, \bar{t}\, b\bar{b} j 
		\to \ell^{+} + \slashed{E}_T + 3b + j + X.
	\end{equation}
	To retain a sufficient number of signal events, we do not attempt to reconstruct the second top quark explicitly and instead allow it to decay inclusively. The charge-conjugate process $pp \to \bar{B}bj$ is also included. The dominant SM backgrounds leading to the same final state are $t\bar{t}b\bar{b}$, $t\bar{t}V$ with $V=h,Z,W$, $t\bar{t}jj$, and $b\bar{b}WW$. Higher-order QCD effects are approximately taken into account through the overall $K$ factors listed in Table~\ref{tab:kfact}.
	
	\begin{table}[t]
		\centering
		\begin{adjustbox}{width=\columnwidth}
			\begin{tabular}{c|cccccc}
				\hline
				Processes  & $t\bar{t}jj$ & $t\bar{t}b\bar{b}$ & $t\bar{t}Z$ & $t\bar{t}W$ & $t\bar{t}h$ & $WWb\bar{b}$\\
				\hline
				$K$ factor  & 1.36~\cite{Bevilacqua:2022ozv} & 1.77~\cite{Bevilacqua:2009zn} & 1.27~\cite{Bevilacqua:2022nrm} & 1.24~\cite{Campbell:2012dh} & 1.21~\cite{Stremmer:2021bnk} & 1.51~\cite{Alwall:2014hca}\\
				\hline
			\end{tabular}
		\end{adjustbox}
		\caption{Overall $K$ factors used for the dominant background processes.}
		\label{tab:kfact}
	\end{table}
	
	\begin{figure*}[t]
		\centering
		\includegraphics[width=\textwidth]{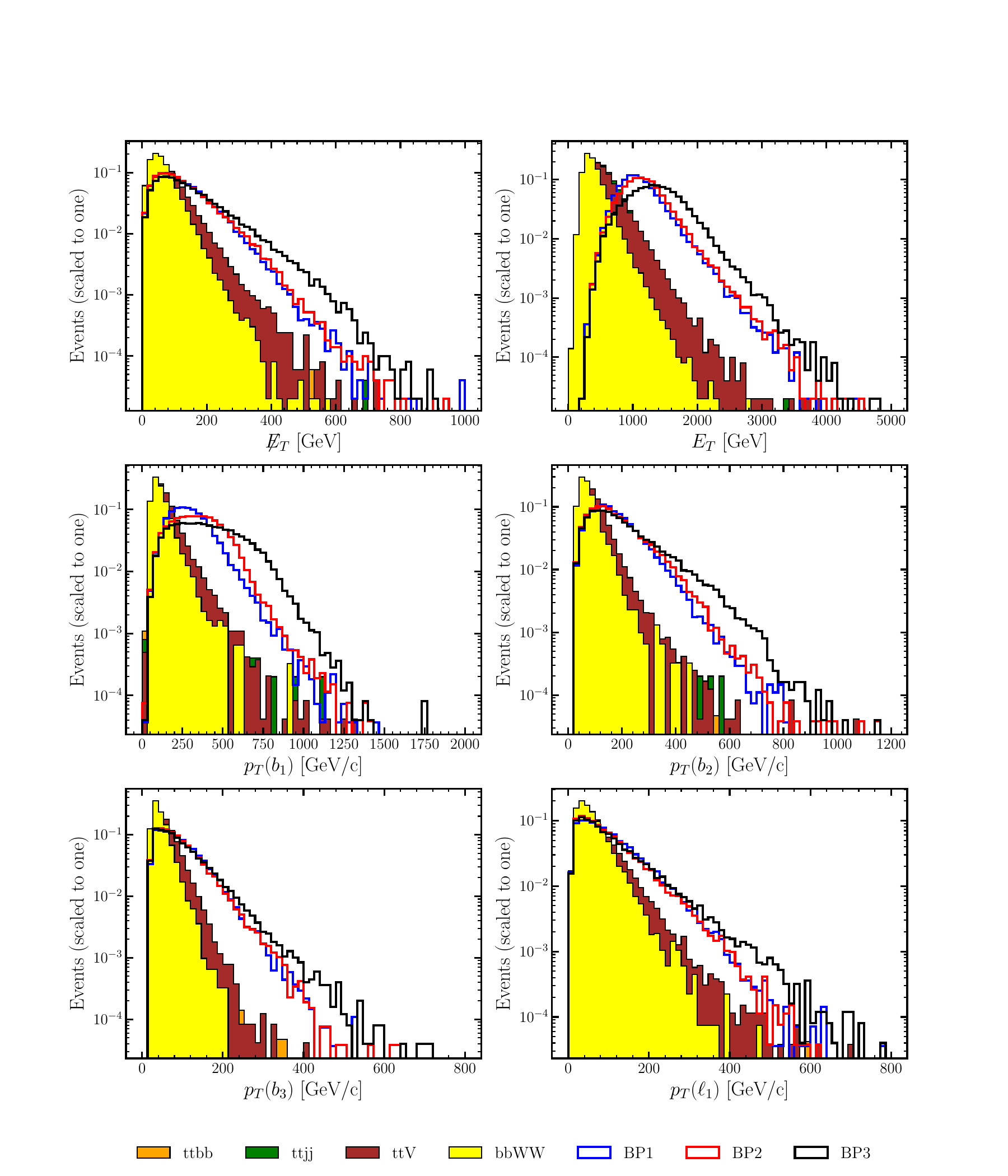}
		\caption{Normalized distributions of representative kinematic observables for the signal and the dominant SM backgrounds at $\sqrt{s}=14$ TeV.}
		\label{fig:obs}
	\end{figure*}
	
	Signal and background samples are generated at parton level using \texttt{MadGraph5\_aMC@NLO} v2.9.14 \cite{Alwall:2014hca}. The generated events are interfaced with \texttt{Pythia 8.30} \cite{Sjostrand:2014zea} for parton showering and hadronization, while detector effects are simulated with \texttt{Delphes 3.5.0} \cite{deFavereau:2013fsa} using the default CMS detector card.
	
	Jets are reconstructed with the anti-$k_t$ algorithm \cite{Cacciari:2008gp} with radius parameter $R=0.4$. The final event selection and statistical analysis are performed within \texttt{MadAnalysis5} \cite{Conte:2013mea}.
	
	At parton level, we impose the following basic cuts on both the signal and the backgrounds:
	$
	\begin{array}{ll}
		p_T^{\ell} > 10~\text{GeV}, & p_T^{j/b}> 20~\text{GeV}, \\
		|\eta_{\ell}| < 2.5, & |\eta_{j/b}| < 5, \\
		\Delta R(x,y) > 0.4, & \forall\, x,y = j,b,\ell.
	\end{array}
	$
	Here, $p_T^{\ell,b,j}$ and $|\eta^{\ell,b,j}|$ denote the transverse momenta and pseudorapidities of leptons, $b$-jets, and light jets, respectively.
	
	In Fig.~\ref{fig:obs}, we display the distributions of several relevant observables for both the signal and the backgrounds, including the missing transverse energy $\slashed{E}_T$, the total transverse energy $E_T$, the lepton transverse momentum $p_T(\ell)$, and the transverse momenta of the three leading $b$-jets, $p_T(b_i)$ with $i=1,2,3$. Guided by these distributions, we define the cut-based event selection summarized in Table~\ref{tab:cuts}.
	
	\begin{table}[t]
		\centering
		\begin{tabular}{lc}
			\toprule
			\textbf{Cuts} & \textbf{Definition} \\
			\midrule
			Cut 1 & $\slashed{E}_T \geq 10$ GeV, $E_T \geq 1100$ GeV \\
			Cut 2 & $p_T(b_1) \geq 450$ GeV, $p_T(b_2) \geq 110$ GeV \\
			Cut 3 & $p_T(b_3) \geq 110$ GeV, $p_T(\ell_1) \geq 10$ GeV \\
			\bottomrule
		\end{tabular}
		\caption{Selection criteria adopted in the cut-based analysis.}
		\label{tab:cuts}
	\end{table}
	
	The cut-flow of the signal and background cross sections, expressed in fb, is shown in Table~\ref{tab:cutflow} for the three benchmark points. The results show that the chosen cuts suppress the SM backgrounds efficiently while preserving a visible signal.
	
	\begin{table}[t]
		\centering
		\begin{adjustbox}{width=\columnwidth}
			\begin{tabular}{|c|ccc|cccc|}
				\hline
				\multirow{2}{*}{Cuts} 
				& \multicolumn{3}{c|}{Signal} 
				& \multicolumn{4}{c|}{Backgrounds} \\
				\cline{2-8}
				& BP1 & BP2 & BP3 & $ttbb$ & $ttV$ & $ttjj$ & $bbWW$ \\
				\hline
				Basic & 2.98 & 2.226 & 0.945 & 298.759 & 125.38 & 69331.01 & 92523.86 \\
				Cut 1 & 1.35 & 1.23 & 0.681 & 7.42 & 5.07 & 2301.8 & 597.7 \\
				Cut 2 & 0.189 & 0.261 & 0.187 & 0.615 & 0.56 & 65.17 & 38.86 \\
				Cut 3 & 0.0437 & 0.0514 & 0.0359 & 0.0928 & 0.075 & 0.00 & 0.00 \\
				\hline
				Eff. [\%] & 1.47 & 2.32 & 3.8 & 0.032 & 0.06 & 0.0 & 0.0 \\
				\hline
			\end{tabular}
		\end{adjustbox}
		\caption{Cut-flow of the cross sections, in fb, for the signal and the dominant SM backgrounds.}
		\label{tab:cutflow}
	\end{table}
	
	\section{Machine-learning analysis}
	\label{sec:ML}
	
\begin{figure*}[htpb!]
		\centering
		\includegraphics[width=\textwidth]{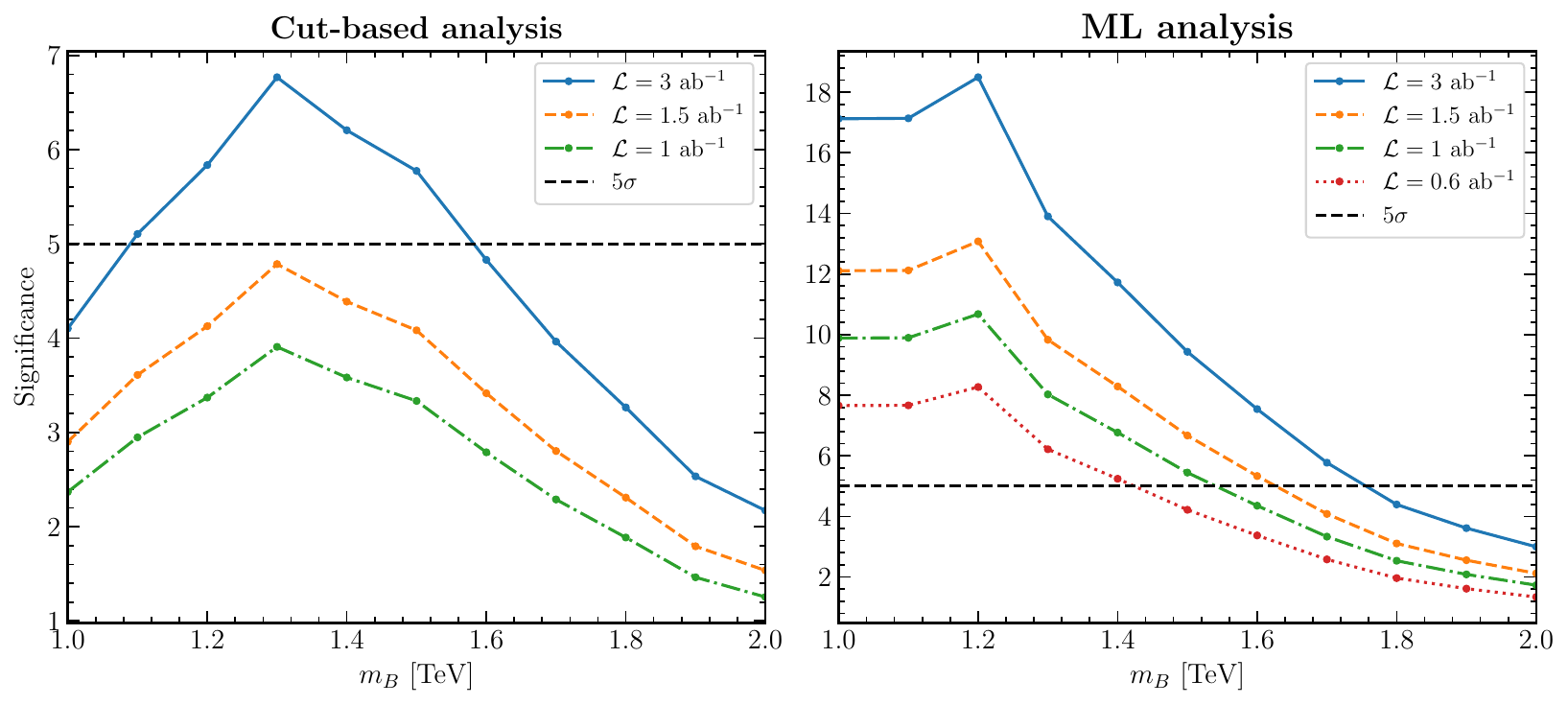}
		\caption{Signal significance as a function of the vector-like bottom mass for integrated luminosities of $\mathcal{L}=3$, $1.5$, $1$, and $0.6~\mathrm{ab}^{-1}$. The horizontal black line indicates the $5\sigma$ discovery threshold. The parameters are those of BP1.}
		\label{fig:sign}
	\end{figure*}

	The cut-based analysis achieves a discovery significance exceeding $5\sigma$ only at very high integrated luminosities, $\mathcal{L} \approx 3~\mathrm{ab}^{-1}$ (see the left panel of Fig.~\ref{fig:sign}). More importantly, this significance is significantly reduced once systematic uncertainties are included. To improve the sensitivity, we therefore employ a multivariate analysis based on boosted decision trees. In particular, we use the XGBoost implementation of extreme gradient boosting \cite{Chen:2016btl}, which provides a powerful discrimination between signal and background. The kinematic variables used as input features are listed in Table~\ref{tab:inputs}. The hyperparameters are chosen as follows: \texttt{n\_estimators}=800, \texttt{max\_depth}=4, and learning rate $=0.02$.
	
	\begin{table}[t]
		\centering\begin{adjustbox}{width=\columnwidth}
		\begin{tabular}{|c|c|}
			\hline
			\textbf{Variable} & \textbf{Description} \\
			\hline
			$P_T(j)$ & Jet transverse momentum \\
			\hline
			$P_T(\ell)$ & Lepton transverse momentum \\
			\hline
			$P_T(b)$ & $b$-jet transverse momentum \\
			\hline
			$\max(\eta_j)$ & Maximum absolute pseudorapidity of jets \\
			\hline
			$M_{\text{inv}}(j_1,j_2)$ & Invariant mass of the leading and subleading jets \\
			\hline
			$M_T(\ell,\slashed{E}_T)$ & Transverse mass of the $W$ boson \\
			\hline
			$\slashed{E}_T$ & Missing transverse energy \\
			\hline
			$N_b$ & Number of $b$-tagged jets \\
			\hline
			$N_{\ell}$ & Number of leptons \\
			\hline
			$N_j$ & Number of jets \\
			\hline
			$\Delta R(j_1,j_2)$ & Angular separation between the two leading jets \\
			\hline
			$\Delta R(b_1,b_2)$ & Angular separation between the two leading $b$-jets \\
			\hline
			$\Delta R(\ell_1,b_1)$ & Angular separation between the leading lepton and leading $b$-jet \\
			\hline
		\end{tabular}		\end{adjustbox}
		\caption{Input variables used in the training and testing of the XGBoost classifier.}
		\label{tab:inputs}
	\end{table}
	
\begin{figure*}[t]
		\centering
		\includegraphics[width=\textwidth]{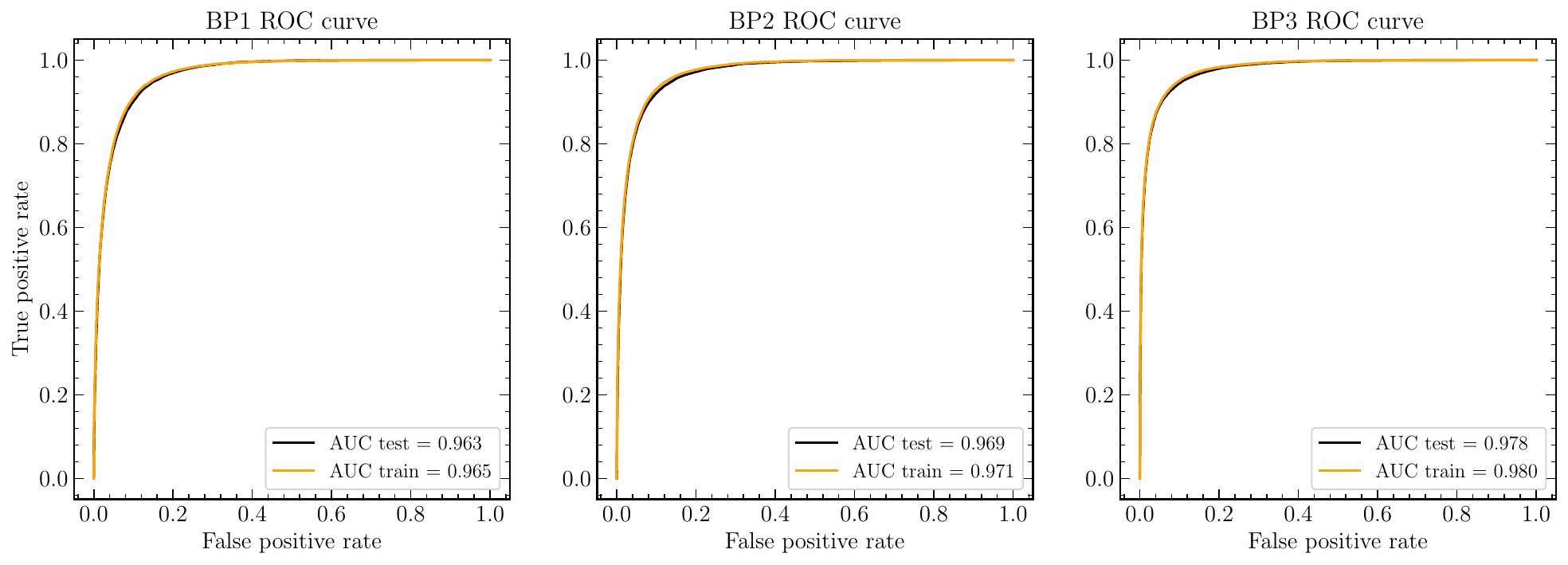}
		\caption{ROC curves for BP1 (left), BP2 (center), and BP3 (right), showing the classifier performance on the training sample (yellow) and test sample (black).}
		\label{fig:roc}
	\end{figure*}

	The performance of the classifier is quantified using the receiver operating characteristic (ROC) curves shown in Fig.~\ref{fig:roc}. The ROC curve illustrates the trade-off between the true positive rate and the false positive rate as the discrimination threshold is varied. The area under the curve (AUC) provides a compact measure of the classifier performance, with AUC$=0.5$ corresponding to random classification and AUC$=1$ to perfect separation.
	
	For BP1, the classifier achieves an AUC of $0.965$ on the training set and $0.963$ on the test set. For BP2, the training and test AUC values are $0.971$ and $0.969$, respectively, while for BP3, the corresponding values are $0.980$ and $0.978$. These results demonstrate a strong separation power between signal and background based on the selected observables. The small differences between the training and test performance also indicate good generalization and no significant evidence of severe overfitting.
	
	To further assess possible overtraining, we compare the training and test samples using the Kolmogorov-Smirnov (KS) test \cite{ross2014introduction}. The KS statistic measures the maximal difference between the cumulative distribution functions of two independent samples,
	\begin{align}
		D = \max_x \left|F_1(x)-F_2(x)\right|,
	\end{align}
	where $F_1(x)$ and $F_2(x)$ denote the corresponding cumulative distributions. The test also provides a $p$-value, which quantifies the probability of obtaining a KS statistic at least as large as the observed one under the null hypothesis that the two samples are drawn from the same parent distribution. A small $p$-value therefore signals a statistically significant discrepancy.
	
\begin{figure*}[t]
		\centering
		\includegraphics[width=\textwidth]{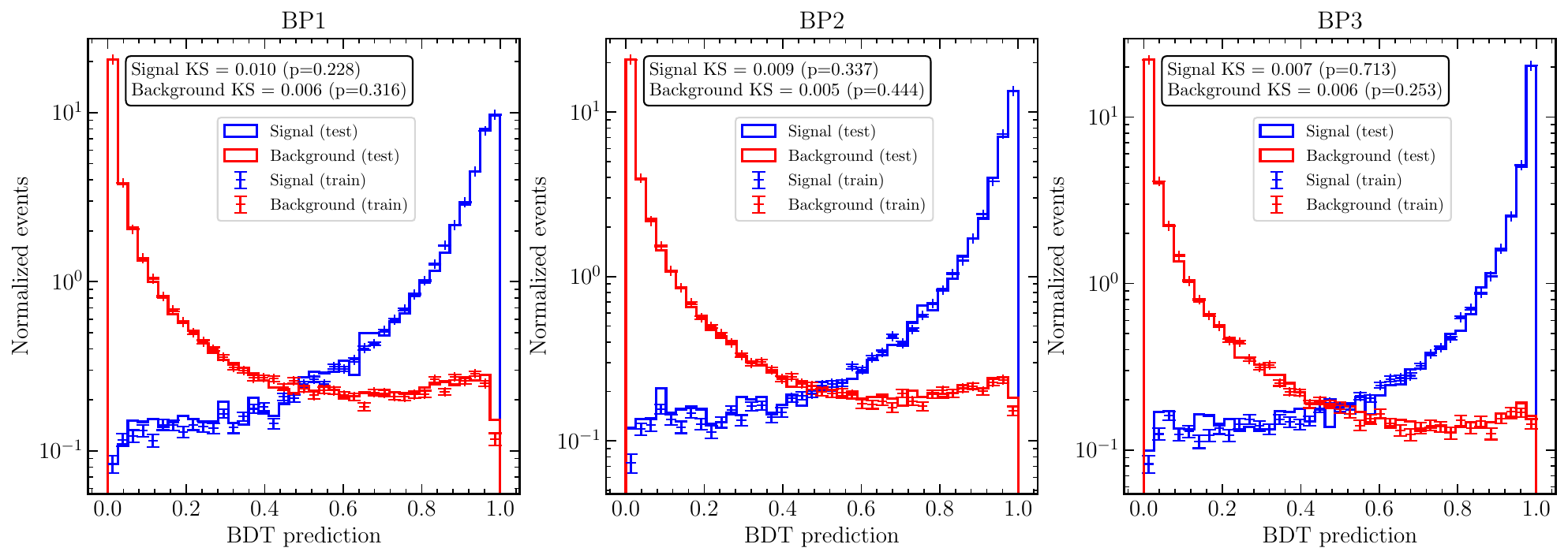}
		\caption{BDT response distributions for the three benchmark points, comparing signal (blue) and background (red) in the training and testing samples. The legend shows the KS statistic and the corresponding $p$-value.}
		\label{fig:ks}
	\end{figure*}

	If $D$ is large and $p<0.05$, the two distributions are significantly different, indicating possible overtraining. Conversely, if $D$ is small or $p\ge 0.05$, the training and test samples are statistically compatible. In Fig.~\ref{fig:ks}, we show the classifier outputs for signal and background in both samples, together with the corresponding KS statistics and $p$-values. In all cases, the KS statistic remains small and one finds $p>5\%$, indicating no significant evidence for overfitting.
	
\begin{figure}[htpb!]
		\centering
		\includegraphics[width=0.45\textwidth]{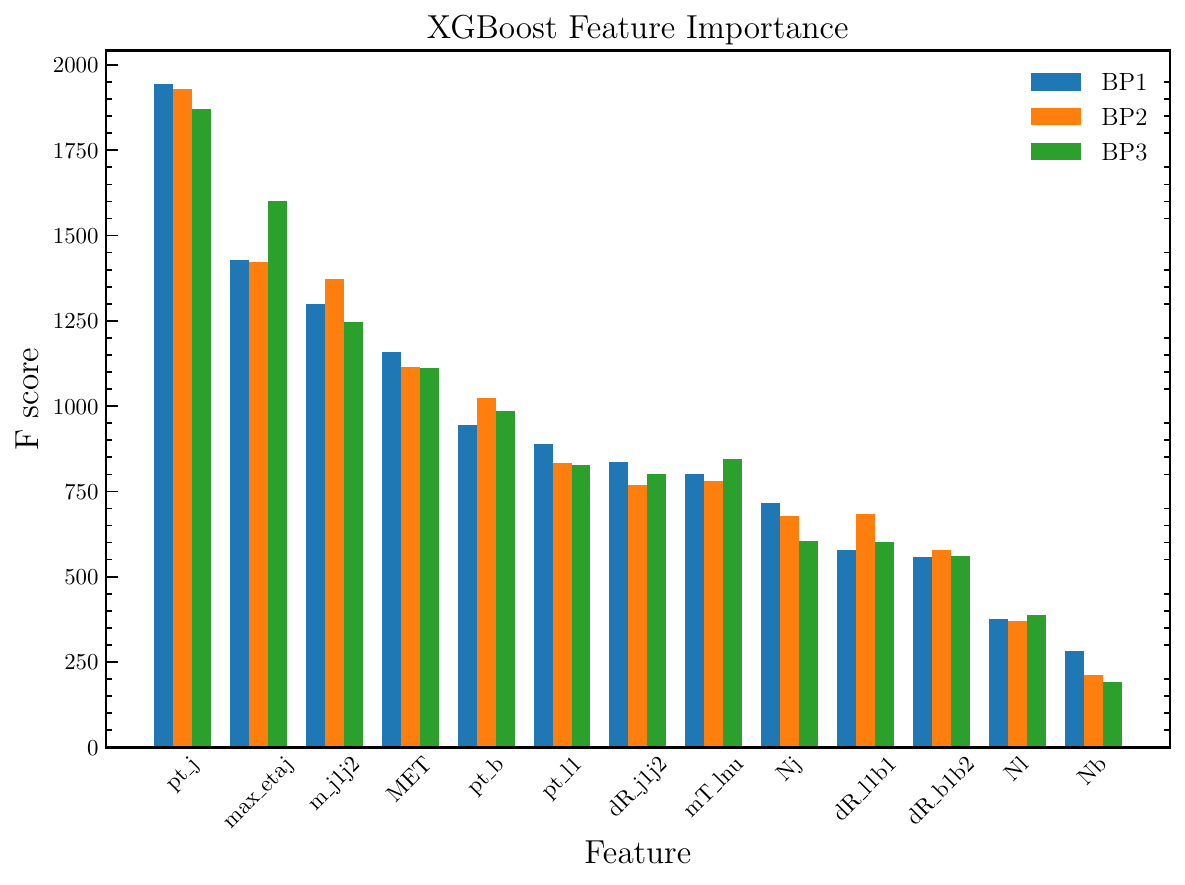}
		\caption{Feature importances extracted from the XGBoost classifier, indicating the variables that contribute most strongly to the signal-background discrimination.}
		\label{fig:features}
	\end{figure}

	Figure~\ref{fig:features} shows the feature importances for BP1, BP2, and BP3. We find that the transverse momentum of the jet, $p_T(j)$, provides the dominant contribution to the signal-background separation, followed by $\max(\eta_j)$, the dijet invariant mass $m_{j_1j_2}$, and the transverse momentum of the $b$-jets, $p_T(b)$. Additional relevant observables include the missing transverse energy, $\slashed{E}_T$, the angular separation $\Delta R(j_1,j_2)$, the leading-lepton transverse momentum $p_T(\ell_1)$, and the transverse mass $M_T(\ell,\slashed{E}_T)$.
	
	The signal significance is first evaluated using the Asimov approximation \cite{Cowan:2010js} in the absence of systematic uncertainty on the background,
	\begin{equation}
		Z_A = \sqrt{2\left[(s+b)\ln\left(1+\frac{s}{b}\right)-s\right]},
	\end{equation}
	where $s$ and $b$ denote the expected numbers of signal and background events, respectively. Including a fractional systematic uncertainty $\delta$ on the background, the significance becomes
	\begin{align}
		\mathcal{Z} = \sqrt{2\left[(s+b)\ln\left(\frac{(s+b)(1+\delta^2 b)}{b+\delta^2 b (s+b)}\right)
			-\frac{1}{\delta^2}\ln\left(1+\delta^2\frac{s}{1+\delta^2 b}\right)\right]}.
	\end{align}
	
	The discovery sensitivity is obtained by scanning the BDT response and selecting the threshold that maximizes $\mathcal{Z}$. Table~\ref{tab:cutvsxg} compares the significances obtained in the cut-based and XGBoost analyses. The improvement brought by the gradient-boosting approach is substantial for all benchmark points.
	
	\begin{table}[htpb!]
		\centering
		\begin{tabular}{lcc}
			\toprule
			\textbf{BPs} & \textbf{Cut-based} & \textbf{XGBoost} \\
			\midrule
			BP1 & 5.02 & 17.81 \\
			BP2 & 5.87 & 13.0 \\
			BP3 & 4.14 & 8.84 \\
			\bottomrule
		\end{tabular}
		\caption{Comparison of the signal significance obtained in the cut-based and XGBoost analyses for the selected benchmark points at $\sqrt{s}=14$ TeV and $\mathcal{L}=3~\mathrm{ab}^{-1}$.}
		\label{tab:cutvsxg}
	\end{table}
	
	We then evaluate the significance for three integrated luminosities, $\mathcal{L}=600,\ 1500,\ 3000~\mathrm{fb}^{-1}$, and for three values of the background systematic uncertainty, $\delta=5\%,\ 10\%,\ 15\%$, as summarized in Table~\ref{tab:mass_lumi}.
	
	The machine-learning analysis yields a substantial improvement in sensitivity over the cut-based strategy. As a consequence, the $5\sigma$ discovery criterion is satisfied over a broad range of vector-like bottom masses. For an integrated luminosity of $\mathcal{L}=600~\mathrm{fb}^{-1}$, the discovery reach extends up to about $m_B \simeq 1.3~\mathrm{TeV}$ for all considered values of the background uncertainty.
	
	With increasing luminosity, the sensitivity improves significantly. In particular, for $\mathcal{L}=1.5~\mathrm{ab}^{-1}$ and $\mathcal{L}=3~\mathrm{ab}^{-1}$, the $5\sigma$ discovery threshold can be reached up to about $m_B \simeq 1.6~\mathrm{TeV}$ when $\delta=5\%$, and up to about $m_B \simeq 1.5~\mathrm{TeV}$ when $\delta=10\%$ and $\delta=15\%$.
	
	\begin{table*}[t]
		\centering
		\begin{tabular}{|c|c|c|c|c|c|c|c|c|c|}
			\hline
			\multirow{2}{*}{\textbf{$m_B$ [TeV]}} 
			& \multicolumn{3}{c|}{\textbf{600 fb$^{-1}$}} 
			& \multicolumn{3}{c|}{\textbf{1500 fb$^{-1}$}} 
			& \multicolumn{3}{c|}{\textbf{3000 fb$^{-1}$}} \\
			\cline{2-10}
			& $\delta=5\%$ & $\delta=10\%$ & $\delta=15\%$
			& $\delta=5\%$ & $\delta=10\%$ & $\delta=15\%$
			& $\delta=5\%$ & $\delta=10\%$ & $\delta=15\%$ \\
			\hline
			1.0 & 7.56 & 7.28 & 6.88 & 11.72 & 10.78 & 9.66 & 16.08 & 13.93 & 11.79 \\
			\hline
			1.1 & 7.56 & 7.28 & 6.89 & 11.73 & 10.78 & 9.66 & 16.10 & 13.94 & 11.80 \\
			\hline
			1.2 & 8.16 & 7.84 & 7.41 & 12.64 & 11.59 & 10.36 & 17.33 & 14.95 & 12.63 \\
			\hline
			1.3 & 6.14 & 5.94 & 5.64 & 9.54 & 8.84 & 7.97 & 13.13 & 11.48 & 9.78 \\
			\hline
			1.4 & 5.18 & 5.02 & 4.79 & 8.07 & 7.50 & 6.80 & 11.12 & 9.79 & 8.38 \\
			\hline
			1.5 & 4.18 & 4.06 & 3.88 & 6.51 & 6.08 & 5.54 & 8.99 & 7.97 & 6.87 \\
			\hline
			1.6 & 3.34 & 3.25 & 3.12 & 5.21 & 4.89 & 4.48 & 7.21 & 6.44 & 5.58 \\
			\hline
			1.7 & 2.56 & 2.50 & 2.40 & 4.00 & 3.77 & 3.46 & 5.54 & 4.98 & 4.34 \\
			\hline
			1.8 & 1.95 & 1.90 & 1.84 & 3.05 & 2.88 & 2.66 & 4.23 & 3.82 & 3.35 \\
			\hline
			1.9 & 1.60 & 1.57 & 1.51 & 2.51 & 2.38 & 2.20 & 3.48 & 3.16 & 2.77 \\
			\hline
			2.0 & 1.33 & 1.30 & 1.26 & 2.08 & 1.98 & 1.83 & 2.89 & 2.63 & 2.32 \\
			\hline
		\end{tabular}
		\caption{Signal significance as a function of the vector-like bottom mass and integrated luminosity for different assumptions on the background systematic uncertainty. The parameters are those of BP1.}
		\label{tab:mass_lumi}
	\end{table*}
	
	Scanning over $m_B$ in the range $1$-$2~\mathrm{TeV}$, we obtain the significance shown in Fig.~\ref{fig:sign} for integrated luminosities of $\mathcal{L}=3~\mathrm{ab}^{-1}$, $1.5~\mathrm{ab}^{-1}$, $1~\mathrm{ab}^{-1}$, and $0.6~\mathrm{ab}^{-1}$, for both the cut-based and the machine-learning analyses. In both approaches, the significance decreases as $m_B$ increases, reflecting the reduction of the signal cross section at larger masses.
	
	A comparison between the two analyses clearly shows that the machine-learning strategy substantially outperforms the conventional cut-based one over the full mass interval. While the cut-based analysis yields only moderate sensitivity and rapidly loses discovery power as the heavy-quark mass increases, the XGBoost analysis consistently delivers much larger significances for all luminosities.
	
	In the cut-based case, the significance reaches only modest values at $\mathcal{L}=3~\mathrm{ab}^{-1}$ and drops below the $5\sigma$ threshold as the mass increases. By contrast, the machine-learning analysis reaches a significance of about $18.5$ at $m_B = 1.2~\mathrm{TeV}$ for $\mathcal{L}=3~\mathrm{ab}^{-1}$ and preserves a significance above the $5\sigma$ level over a substantially wider mass interval. This highlights the superior capability of multivariate methods in isolating the signal from the large SM background.
	
\section{Conclusion}
\label{sec:Conclusions}

We have analyzed the discovery potential of a singly produced vector-like bottom quark in the Type-II 2HDM extended by an $SU(2)_L$ $(T,B)$ vector-like quark doublet, focusing on the non-standard decay chain $B \to \phi b$ with $\phi = H, A$ and $\phi \to t\bar{t}$. In the parameter region considered in this work, the exotic decays into heavy neutral Higgs bosons can dominate over the conventional SM channels, with $\mathrm{BR}(B \to Hb)$ and $\mathrm{BR}(B \to Ab)$ reaching values close to $50\%$ in the alignment limit.

After imposing the relevant theoretical and experimental constraints, we identified viable benchmark configurations for which the signal process
$
pp \to B\bar{b}j \to \phi b\bar{b}j \to t\bar{t}b\bar{b}j
\to W^{+} b\, \bar{t}\, b\bar{b} j \to \ell^{+} + \slashed{E}_T + 3b + j + X
$
can be studied at the HL-LHC. Our cut-based analysis shows that the chosen kinematic selection suppresses the dominant SM backgrounds efficiently, but the resulting sensitivity remains limited once systematic uncertainties are included.

To improve the reach, we performed a multivariate analysis based on XGBoost. The classifier exhibits strong discriminating power, with high AUC values and no significant indication of overtraining according to the Kolmogorov-Smirnov test. Compared with the cut-based analysis, the multivariate approach leads to a substantial enhancement in the signal significance for all benchmark points considered.

For the benchmark scenarios studied here, the XGBoost analysis gives a $5\sigma$ discovery reach up to about $m_B \simeq 1.3~\mathrm{TeV}$ with $600~\mathrm{fb}^{-1}$ and up to about $m_B \simeq 1.6~\mathrm{TeV}$ with $3~\mathrm{ab}^{-1}$, even after including background systematic uncertainties of up to $15\%$. Our results therefore show that this channel can be relevant at the HL-LHC and can complement existing searches for heavy neutral Higgs bosons and vector-like quarks.
	
	\begin{acknowledgments}
		M.~Boukidi acknowledges the support of Narodowe Centrum Nauki under OPUS Grant No.~2023/49/B/ST2/03862. The authors are grateful for the
technical support provided by CNRST/HPC-MARWAN.
	\end{acknowledgments}
	
	\bibliographystyle{apsrev4-2}
	\bibliography{main.bib}

@article{ParticleDataGroup:2020ssz,
    author = "Zyla, P. A. and others",
    collaboration = "Particle Data Group",
    title = "{Review of Particle Physics}",
    doi = "10.1093/ptep/ptaa104",
    journal = "PTEP",
    volume = "2020",
    number = "8",
    pages = "083C01",
    year = "2020"
}

@book{Gunion:1989we,
    author = "Gunion, John F. and Haber, Howard E. and Kane, Gordon L. and Dawson, Sally",
    title = "{The Higgs Hunter's Guide}",
    reportNumber = "SCIPP-89/13, UCD-89-4, BNL-41644",
    doi = "10.1201/9780429496448",
    isbn = "978-0-429-49644-8",
    volume = "80",
    year = "2000"
}

@book{Moretti:2019ulc,
    author = "Moretti, Stefano and Khalil, Shaaban",
    title = "{Supersymmetry Beyond Minimality: From Theory to Experiment}",
    isbn = "978-0-367-87662-3",
    publisher = "CRC Press",
    year = "2019"
}

@article{Okada:2012gy,
    author = "Okada, Yasuhiro and Panizzi, Luca",
    title = "{LHC signatures of vector-like quarks}",
    eprint = "1207.5607",
    archivePrefix = "arXiv",
    primaryClass = "hep-ph",
    reportNumber = "KEK-TH-1560, LYCEN-2012-04",
    doi = "10.1155/2013/364936",
    journal = "Adv. High Energy Phys.",
    volume = "2013",
    pages = "364936",
    year = "2013"
}

@article{DeSimone:2012fs,
    author = "De Simone, Andrea and Matsedonskyi, Oleksii and Rattazzi, Riccardo and Wulzer, Andrea",
    title = "{A First Top Partner Hunter's Guide}",
    eprint = "1211.5663",
    archivePrefix = "arXiv",
    primaryClass = "hep-ph",
    reportNumber = "CERN-PH-TH-2012-323, SISSA-31-2012-EP",
    doi = "10.1007/JHEP04(2013)004",
    journal = "JHEP",
    volume = "04",
    pages = "004",
    year = "2013"
}

@article{Cowan:2010js,
    author = "Cowan, Glen and Cranmer, Kyle and Gross, Eilam and Vitells, Ofer",
    title = "{Asymptotic formulae for likelihood-based tests of new physics}",
    eprint = "1007.1727",
    archivePrefix = "arXiv",
    primaryClass = "physics.data-an",
    doi = "10.1140/epjc/s10052-011-1554-0",
    journal = "Eur. Phys. J. C",
    volume = "71",
    pages = "1554",
    year = "2011",
    note = "[Erratum: Eur.Phys.J.C 73, 2501 (2013)]"
}

@article{Cao:2022mif,
    author = "Cao, Junjie and Meng, Lei and Shang, Liangliang and Wang, Shiyu and Yang, Bingfang",
    title = "{Interpreting the W-mass anomaly in vectorlike quark models}",
    eprint = "2204.09477",
    archivePrefix = "arXiv",
    primaryClass = "hep-ph",
    doi = "10.1103/PhysRevD.106.055042",
    journal = "Phys. Rev. D",
    volume = "106",
    number = "5",
    pages = "055042",
    year = "2022"
}

@article{Eriksson:2009ws,
    author = "Eriksson, David and Rathsman, Johan and Stal, Oscar",
    title = "{2HDMC: Two-Higgs-Doublet Model Calculator Physics and Manual}",
    eprint = "0902.0851",
    archivePrefix = "arXiv",
    primaryClass = "hep-ph",
    doi = "10.1016/j.cpc.2009.09.011",
    journal = "Comput. Phys. Commun.",
    volume = "181",
    pages = "189--205",
    year = "2010"
}

@article{Alwall:2014hca,
    author = "Alwall, J. and Frederix, R. and Frixione, S. and Hirschi, V. and Maltoni, F. and Mattelaer, O. and Shao, H. -S. and Stelzer, T. and Torrielli, P. and Zaro, M.",
    title = "{The automated computation of tree-level and next-to-leading order differential cross sections, and their matching to parton shower simulations}",
    eprint = "1405.0301",
    archivePrefix = "arXiv",
    primaryClass = "hep-ph",
    reportNumber = "CERN-PH-TH-2014-064, CP3-14-18, LPN14-066, MCNET-14-09, ZU-TH-14-14",
    doi = "10.1007/JHEP07(2014)079",
    journal = "JHEP",
    volume = "07",
    pages = "079",
    year = "2014"
}

@article{Bechtle:2020uwn,
    author = "Bechtle, Philip and Heinemeyer, Sven and Klingl, Tobias and Stefaniak, Tim and Weiglein, Georg and Wittbrodt, Jonas",
    title = "{HiggsSignals-2: Probing new physics with precision Higgs measurements in the LHC 13 TeV era}",
    eprint = "2012.09197",
    archivePrefix = "arXiv",
    primaryClass = "hep-ph",
    reportNumber = "BONN-TH-2020-09, DESY-20-228, DESY 20-228, IFT-UAM/CSIC-20-081, LU TP 20-53",
    doi = "10.1140/epjc/s10052-021-08942-y",
    journal = "Eur. Phys. J. C",
    volume = "81",
    number = "2",
    pages = "145",
    year = "2021"
}

@article{Bechtle:2008jh,
    author = "Bechtle, Philip and Brein, Oliver and Heinemeyer, Sven and Weiglein, Georg and Williams, Karina E.",
    title = "{HiggsBounds: Confronting Arbitrary Higgs Sectors with Exclusion Bounds from LEP and the Tevatron}",
    eprint = "0811.4169",
    archivePrefix = "arXiv",
    primaryClass = "hep-ph",
    reportNumber = "DCPT-08-172, IPPP-08-86, BONN-TH-2008-17",
    doi = "10.1016/j.cpc.2009.09.003",
    journal = "Comput. Phys. Commun.",
    volume = "181",
    pages = "138--167",
    year = "2010"
}

@article{Bechtle:2011sb,
    author = "Bechtle, Philip and Brein, Oliver and Heinemeyer, Sven and Weiglein, Georg and Williams, Karina E.",
    title = "{HiggsBounds 2.0.0: Confronting Neutral and Charged Higgs Sector Predictions with Exclusion Bounds from LEP and the Tevatron}",
    eprint = "1102.1898",
    archivePrefix = "arXiv",
    primaryClass = "hep-ph",
    reportNumber = "FR-PHENO-2011-002, BONN-TH-2011-02, DESY-11-016",
    doi = "10.1016/j.cpc.2011.07.015",
    journal = "Comput. Phys. Commun.",
    volume = "182",
    pages = "2605--2631",
    year = "2011"
}

@article{Bechtle:2013wla,
    author = "Bechtle, Philip and Brein, Oliver and Heinemeyer, Sven and St\r{a}l, Oscar and Stefaniak, Tim and Weiglein, Georg and Williams, Karina E.",
    title = "{$\mathsf{HiggsBounds}-4$: Improved Tests of Extended Higgs Sectors against Exclusion Bounds from LEP, the Tevatron and the LHC}",
    eprint = "1311.0055",
    archivePrefix = "arXiv",
    primaryClass = "hep-ph",
    reportNumber = "BONN-TH-2013-21, DESY-13-110",
    doi = "10.1140/epjc/s10052-013-2693-2",
    journal = "Eur. Phys. J. C",
    volume = "74",
    number = "3",
    pages = "2693",
    year = "2014"
}

@article{Bechtle:2015pma,
    author = "Bechtle, Philip and Heinemeyer, Sven and Stal, Oscar and Stefaniak, Tim and Weiglein, Georg",
    title = "{Applying Exclusion Likelihoods from LHC Searches to Extended Higgs Sectors}",
    eprint = "1507.06706",
    archivePrefix = "arXiv",
    primaryClass = "hep-ph",
    reportNumber = "BONN-TH-2015-08, DESY-15-093, SCIPP-15-05",
    doi = "10.1140/epjc/s10052-015-3650-z",
    journal = "Eur. Phys. J. C",
    volume = "75",
    number = "9",
    pages = "421",
    year = "2015"
}

@article{Arhrib:2024dou,
    author = "Arhrib, Abdesslam and Benbrik, Rachid and Boukidi, Mohammed and Moretti, Stefano",
    title = "{Anatomy of vector-like bottom-quark models in the alignment limit of the 2-Higgs doublet model type-II}",
    eprint = "2403.13021",
    archivePrefix = "arXiv",
    primaryClass = "hep-ph",
    doi = "10.1140/epjc/s10052-024-13390-5",
    journal = "Eur. Phys. J. C",
    volume = "84",
    number = "10",
    pages = "1008",
    year = "2024"
}

@article{Arhrib:2024mbq,
	author = "Arhrib, Abdesslam and Benbrik, Rachid and Boukidi, Mohammed and Moretti, Stefano",
	title = "{Large hadron collider signatures of exotic vector-like quarks within the 2-Higgs doublet model type-II}",
	eprint = "2409.20104",
	archivePrefix = "arXiv",
	primaryClass = "hep-ph",
	doi = "10.1088/1361-6471/ae09be",
	journal = "J. Phys. G",
	volume = "52",
	number = "10",
	pages = "105002",
	year = "2025"
}

@article{Buchkremer:2013bha,
    author = "Buchkremer, Mathieu and Cacciapaglia, Giacomo and Deandrea, Aldo and Panizzi, Luca",
    title = "{Model Independent Framework for Searches of Top Partners}",
    eprint = "1305.4172",
    archivePrefix = "arXiv",
    primaryClass = "hep-ph",
    reportNumber = "LYCEN-2013-03, SHEP-13-10, CP3-13-22",
    doi = "10.1016/j.nuclphysb.2013.08.010",
    journal = "Nucl. Phys. B",
    volume = "876",
    pages = "376--417",
    year = "2013"
}

@article{Han:2025itd,
	author = "Han, Jin-Zhong and Liu, Yao-Bei and Moretti, Stefano",
	title = "{Searching for single production of vectorlike quarks decaying into Wb at a future muon-proton collider}",
	eprint = "2501.01026",
	archivePrefix = "arXiv",
	primaryClass = "hep-ph",
	doi = "10.1103/6njz-xvr6",
	journal = "Phys. Rev. D",
	volume = "112",
	number = "3",
	pages = "035016",
	year = "2025"
}

@article{Arkani-Hamed:2002iiv,
    author = "Arkani-Hamed, N. and Cohen, A. G. and Katz, E. and Nelson, A. E. and Gregoire, T. and Wacker, Jay G.",
    title = "{The Minimal moose for a little Higgs}",
    eprint = "hep-ph/0206020",
    archivePrefix = "arXiv",
    reportNumber = "BUHEP-02-24, UW-PT-01-09, HUTP-02-A016",
    doi = "10.1088/1126-6708/2002/08/021",
    journal = "JHEP",
    volume = "08",
    pages = "021",
    year = "2002"
}

@article{Han:2003wu,
    author = "Han, Tao and Logan, Heather E. and McElrath, Bob and Wang, Lian-Tao",
    title = "{Phenomenology of the little Higgs model}",
    eprint = "hep-ph/0301040",
    archivePrefix = "arXiv",
    reportNumber = "MADPH-02-1317",
    doi = "10.1103/PhysRevD.67.095004",
    journal = "Phys. Rev. D",
    volume = "67",
    pages = "095004",
    year = "2003"
}

@article{Chang:2003vs,
    author = "Chang, Spencer and He, Hong-Jian",
    title = "{Unitarity of little Higgs models signals new physics of UV completion}",
    eprint = "hep-ph/0311177",
    archivePrefix = "arXiv",
    reportNumber = "HUTP-03-A075, UTHEP-03-19",
    doi = "10.1016/j.physletb.2004.02.027",
    journal = "Phys. Lett. B",
    volume = "586",
    pages = "95--105",
    year = "2004"
}

@article{Yang:2024aav,
    author = "Yang, Bingfang and Li, Zejun and Jia, Xinglong and Moretti, Stefano and Shang, Liangliang",
    title = "{Search for single vector-like B quark production in hadronic final states at the LHC}",
    eprint = "2405.13452",
    archivePrefix = "arXiv",
    primaryClass = "hep-ph",
    doi = "10.1140/epjc/s10052-024-13482-2",
    journal = "Eur. Phys. J. C",
    volume = "84",
    number = "10",
    pages = "1124",
    year = "2024"
}

@article{Contino:2006qr,
    author = "Contino, Roberto and Da Rold, Leandro and Pomarol, Alex",
    title = "{Light custodians in natural composite Higgs models}",
    eprint = "hep-ph/0612048",
    archivePrefix = "arXiv",
    reportNumber = "UAB-FT-619, ROMA1-1445-2006",
    doi = "10.1103/PhysRevD.75.055014",
    journal = "Phys. Rev. D",
    volume = "75",
    pages = "055014",
    year = "2007"
}

@article{Matsedonskyi:2012ym,
    author = "Matsedonskyi, Oleksii and Panico, Giuliano and Wulzer, Andrea",
    title = "{Light Top Partners for a Light Composite Higgs}",
    eprint = "1204.6333",
    archivePrefix = "arXiv",
    primaryClass = "hep-ph",
    doi = "10.1007/JHEP01(2013)164",
    journal = "JHEP",
    volume = "01",
    pages = "164",
    year = "2013"
}

@article{Lodone:2008yy,
    author = "Lodone, Paolo",
    title = "{Vector-like quarks in a 'composite' Higgs model}",
    eprint = "0806.1472",
    archivePrefix = "arXiv",
    primaryClass = "hep-ph",
    doi = "10.1088/1126-6708/2008/12/029",
    journal = "JHEP",
    volume = "12",
    pages = "029",
    year = "2008"
}

@article{Benbrik:2025nfw,
    author = "Benbrik, R. and Berrouj, M. and Boukidi, M. and Kahime, K.",
    title = "{Exploring vector-like top quark pair production via charged Higgs decays in multi-b-jet and opposite-sign dilepton final state at the LHC}",
    doi = "10.1140/epjc/s10052-025-14237-3",
    journal = "Eur. Phys. J. C",
    volume = "85",
    number = "5",
    pages = "500",
    year = "2025"
}

@article{Benbrik:2022kpo,
    author = "Benbrik, Rachid and Boukidi, Mohammed and Moretti, Stefano",
    title = "{Probing charged Higgs bosons in the two-Higgs-doublet model type II with vectorlike quarks}",
    eprint = "2211.07259",
    archivePrefix = "arXiv",
    primaryClass = "hep-ph",
    doi = "10.1103/PhysRevD.109.055016",
    journal = "Phys. Rev. D",
    volume = "109",
    number = "5",
    pages = "055016",
    year = "2024"
}

@article{Abouabid:2023mbu,
    author = "Abouabid, Hamza and Arhrib, Abdesslam and Benbrik, Rachid and Boukidi, Mohammed and Falaki, Jaouad El",
    title = "{The oblique parameters in the 2HDM with vector-like quarks: confronting M $_{W}$ CDF-II anomaly}",
    eprint = "2302.07149",
    archivePrefix = "arXiv",
    primaryClass = "hep-ph",
    doi = "10.1088/1361-6471/ad3f34",
    journal = "J. Phys. G",
    volume = "51",
    number = "7",
    pages = "075001",
    year = "2024"
}

@article{Benbrik:2024hsf,
    author = "Benbrik, Rachid and Berrouj, Mbark and Boukidi, Mohammed",
    title = "{Investigation of charged Higgs bosons production from vectorlike T quark decays at e{\ensuremath{\gamma}} collider}",
    eprint = "2408.15985",
    archivePrefix = "arXiv",
    primaryClass = "hep-ph",
    doi = "10.1103/PhysRevD.111.015027",
    journal = "Phys. Rev. D",
    volume = "111",
    number = "1",
    pages = "015027",
    year = "2025"
}

@article{Arhrib:2024nbj,
    author = "Arhrib, Abdesslam and Benbrik, Rachid and Berrouj, Mbark and Boukidi, Mohammed and Manaut, Bouzid",
    title = "{Search for charged Higgs bosons through vectorlike top quark pair production at the LHC}",
    eprint = "2407.01348",
    archivePrefix = "arXiv",
    primaryClass = "hep-ph",
    doi = "10.1103/PhysRevD.111.095026",
    journal = "Phys. Rev. D",
    volume = "111",
    number = "9",
    pages = "095026",
    year = "2025"
}

@article{Benbrik:2024bxt,
    author = "Benbrik, Rachid and Boukidi, Mohammed and Moretti, Stefano",
    title = "{Charged Higgs Boson Mass Bounds in 2HDM-II: Impact of Vector-Like Quarks}",
    eprint = "2409.16054",
    archivePrefix = "arXiv",
    primaryClass = "hep-ph",
    doi = "10.22323/1.476.0083",
    journal = "PoS",
    volume = "ICHEP2024",
    pages = "083",
    year = "2025"
}

@article{Dermisek:2019vkc,
    author = "Derm{\'\i}{\v{s}}ek, Radovan and Lunghi, Enrico and Shin, Seodong",
    title = "{Hunting for Vectorlike Quarks}",
    eprint = "1901.03709",
    archivePrefix = "arXiv",
    primaryClass = "hep-ph",
    reportNumber = "EFI-19-1",
    doi = "10.1007/JHEP04(2019)019",
    journal = "JHEP",
    volume = "04",
    pages = "019",
    year = "2019",
    note = "[Erratum: JHEP 10, 058 (2020)]"
}

@article{Ghosh:2023xhs,
    author = "Ghosh, Anupam and Konar, Partha",
    title = "{Precision prediction of a democratic up-family philic KSVZ axion model at the LHC}",
    eprint = "2305.08662",
    archivePrefix = "arXiv",
    primaryClass = "hep-ph",
    doi = "10.1016/j.dark.2024.101746",
    journal = "Phys. Dark Univ.",
    volume = "47",
    pages = "101746",
    year = "2025"
}

@article{Benbrik:2019zdp,
    author = "Benbrik, Rachid and others",
    title = "{Signatures of vector-like top partners decaying into new neutral scalar or pseudoscalar bosons}",
    eprint = "1907.05929",
    archivePrefix = "arXiv",
    primaryClass = "hep-ph",
    doi = "10.1007/JHEP05(2020)028",
    journal = "JHEP",
    volume = "05",
    pages = "028",
    year = "2020"
}

@misc{Gunion:1992hs,
	author        = "...",
	title         = "{Errata for the Higgs hunter's guide}",
	eprint        = "hep-ph/9302272",
	archivePrefix = "arXiv",
	year          = "1992"
}

@article{Branco:2011iw,
    author = "Branco, G. C. and Ferreira, P. M. and Lavoura, L. and Rebelo, M. N. and Sher, Marc and Silva, Joao P.",
    title = "{Theory and phenomenology of two-Higgs-doublet models}",
    eprint = "1106.0034",
    archivePrefix = "arXiv",
    primaryClass = "hep-ph",
    doi = "10.1016/j.physrep.2012.02.002",
    journal = "Phys. Rept.",
    volume = "516",
    pages = "1--102",
    year = "2012"
}

@article{Agashe:2004rs,
    author = "Agashe, Kaustubh and Contino, Roberto and Pomarol, Alex",
    title = "{The Minimal composite Higgs model}",
    eprint = "hep-ph/0412089",
    archivePrefix = "arXiv",
    reportNumber = "UAB-FT-567",
    doi = "10.1016/j.nuclphysb.2005.04.035",
    journal = "Nucl. Phys. B",
    volume = "719",
    pages = "165--187",
    year = "2005"
}

@article{Bellazzini:2014yua,
    author = "Bellazzini, Brando and Cs\'aki, Csaba and Serra, Javi",
    title = "{Composite Higgses}",
    eprint = "1401.2457",
    archivePrefix = "arXiv",
    primaryClass = "hep-ph",
    doi = "10.1140/epjc/s10052-014-2766-x",
    journal = "Eur. Phys. J. C",
    volume = "74",
    number = "5",
    pages = "2766",
    year = "2014"
}

@article{Arhrib:2024tzm,
    author = "Arhrib, Abdesslam and Benbrik, Rachid and Boukidi, Mohammed and Manaut, Bouzid and Moretti, Stefano",
    title = "{Anatomy of vector-like top-quark models in the alignment limit of the 2-Higgs Doublet Model Type-II}",
    eprint = "2401.16219",
    archivePrefix = "arXiv",
    primaryClass = "hep-ph",
    doi = "10.1140/epjc/s10052-024-13692-8",
    journal = "Eur. Phys. J. C",
    volume = "85",
    number = "1",
    pages = "2",
    year = "2025"
}

@article{Hewett:1988xc,
    author = "Hewett, JoAnne L. and Rizzo, Thomas G.",
    title = "{Low-Energy Phenomenology of Superstring Inspired E(6) Models}",
    reportNumber = "MAD-PH-446, IS-J-3005",
    doi = "10.1016/0370-1573(89)90071-9",
    journal = "Phys. Rept.",
    volume = "183",
    pages = "193",
    year = "1989"
}

@article{Dermisek:2020gbr,
    author = "Dermisek, Radovan and Lunghi, Enrico and McGinnis, Navin and Shin, Seodong",
    title = "{Signals with six bottom quarks for charged and neutral Higgs bosons}",
    eprint = "2005.07222",
    archivePrefix = "arXiv",
    primaryClass = "hep-ph",
    doi = "10.1007/JHEP07(2020)241",
    journal = "JHEP",
    volume = "07",
    pages = "241",
    year = "2020"
}

@article{Dermisek:2021zjd,
    author = "Dermisek, Radovan and Lunghi, Enrico and Mcginnis, Navin and Shin, Seodong",
    title = "{Tau-jet signatures of vectorlike quark decays to heavy charged and neutral Higgs bosons}",
    eprint = "2105.10790",
    archivePrefix = "arXiv",
    primaryClass = "hep-ph",
    doi = "10.1007/JHEP08(2021)159",
    journal = "JHEP",
    volume = "08",
    pages = "159",
    year = "2021"
}

@article{ATLAS:2023ixh,
    author = "Aad, Georges and others",
    collaboration = "ATLAS",
    title = "{Search for single vector-like B quark production and decay via B \textrightarrow{} bH($ b\overline{b} $) in pp collisions at $ \sqrt{s} $ = 13 TeV with the ATLAS detector}",
    eprint = "2308.02595",
    archivePrefix = "arXiv",
    primaryClass = "hep-ex",
    reportNumber = "CERN-EP-2023-091",
    doi = "10.1007/JHEP11(2023)168",
    journal = "JHEP",
    volume = "11",
    pages = "168",
    year = "2023"
}

@article{Bechtle:2020pkv,
    author = "Bechtle, Philip and Dercks, Daniel and Heinemeyer, Sven and Klingl, Tobias and Stefaniak, Tim and Weiglein, Georg and Wittbrodt, Jonas",
    title = "{HiggsBounds-5: Testing Higgs Sectors in the LHC 13 TeV Era}",
    eprint = "2006.06007",
    archivePrefix = "arXiv",
    primaryClass = "hep-ph",
    reportNumber = "BONN-TH-2020-03, DESY 20-093, DESY-20-093, IFT-UAM/CSIC-20-072, LU 20-27",
    doi = "10.1140/epjc/s10052-020-08557-9",
    journal = "Eur. Phys. J. C",
    volume = "80",
    number = "12",
    pages = "1211",
    year = "2020"
}

@article{Benbrik:2023xlo,
    author = "Benbrik, R. and Berrouj, M. and Boukidi, M. and Habjia, A. and Ghourmin, E. and Rahili, L.",
    title = "{Search for single production of vector-like top partner T\(\rightarrow\)H$^+$b and H\(^\pm\)\(\rightarrow\)tb\textasciimacron{} at the LHC Run-III}",
    doi = "10.1016/j.physletb.2023.138024",
    journal = "Phys. Lett. B",
    volume = "843",
    pages = "138024",
    year = "2023"
}

@article{Bahl:2022igd,
    author = {Bahl, Henning and Biek\"otter, Thomas and Heinemeyer, Sven and Li, Cheng and Paasch, Steven and Weiglein, Georg and Wittbrodt, Jonas},
    title = "{HiggsTools: BSM scalar phenomenology with new versions of HiggsBounds and HiggsSignals}",
    eprint = "2210.09332",
    archivePrefix = "arXiv",
    primaryClass = "hep-ph",
    doi = "10.1016/j.cpc.2023.108803",
    journal = "Comput. Phys. Commun.",
    volume = "291",
    pages = "108803",
    year = "2023"
}

@article{Benbrik:2024fku,
    author = "Benbrik, Rachid and Boukidi, Mohammed and Ech-chaouy, Mohamed and Moretti, Stefano and Salime, Khawla and Yan, Qi-Shu",
    title = "{Vector-Like Quarks at the LHC: A unified perspective from ATLAS and CMS exclusion limits}",
    eprint = "2412.01761",
    archivePrefix = "arXiv",
    primaryClass = "hep-ph",
    doi = "10.1007/JHEP03(2025)020",
    journal = "JHEP",
    volume = "03",
    pages = "020",
    year = "2025"
}

@article{Bhardwaj:2022nko,
    author = "Bhardwaj, Akanksha and Mandal, Tanumoy and Mitra, Subhadip and Neeraj, Cyrin",
    title = "{Roadmap to explore vectorlike quarks decaying to a new scalar or pseudoscalar}",
    eprint = "2203.13753",
    archivePrefix = "arXiv",
    primaryClass = "hep-ph",
    doi = "10.1103/PhysRevD.106.095014",
    journal = "Phys. Rev. D",
    volume = "106",
    number = "9",
    pages = "095014",
    year = "2022"
}

@article{Kanemura:1993hm,
    author = "Kanemura, Shinya and Kubota, Takahiro and Takasugi, Eiichi",
    title = "{Lee-Quigg-Thacker bounds for Higgs boson masses in a two doublet model}",
    eprint = "hep-ph/9303263",
    archivePrefix = "arXiv",
    reportNumber = "OS-GE-32-93",
    doi = "10.1016/0370-2693(93)91205-2",
    journal = "Phys. Lett. B",
    volume = "313",
    pages = "155--160",
    year = "1993"
}

@article{Arhrib:2016rlj,
    author = "Arhrib, A. and Benbrik, R. and King, S. J. D. and Manaut, B. and Moretti, S. and Un, C. S.",
    title = "{Phenomenology of 2HDM with vectorlike quarks}",
    eprint = "1607.08517",
    archivePrefix = "arXiv",
    primaryClass = "hep-ph",
    doi = "10.1103/PhysRevD.97.095015",
    journal = "Phys. Rev. D",
    volume = "97",
    pages = "095015",
    year = "2018"
}

@article{Barroso:2013awa,
    author = "Barroso, A. and Ferreira, P. M. and Ivanov, I. P. and Santos, Rui",
    title = "{Metastability bounds on the two Higgs doublet model}",
    eprint = "1303.5098",
    archivePrefix = "arXiv",
    primaryClass = "hep-ph",
    doi = "10.1007/JHEP06(2013)045",
    journal = "JHEP",
    volume = "06",
    pages = "045",
    year = "2013"
}

@article{Deshpande:1977rw,
    author = "Deshpande, Nilendra G. and Ma, Ernest",
    title = "{Pattern of Symmetry Breaking with Two Higgs Doublets}",
    reportNumber = "OITS-81",
    doi = "10.1103/PhysRevD.18.2574",
    journal = "Phys. Rev. D",
    volume = "18",
    pages = "2574",
    year = "1978"
}

@article{Gopalakrishna:2015wwa,
    author = "Gopalakrishna, Shrihari and Mukherjee, Tuhin Subhra and Sadhukhan, Soumya",
    title = "{Extra neutral scalars with vectorlike fermions at the LHC}",
    eprint = "1504.01074",
    archivePrefix = "arXiv",
    primaryClass = "hep-ph",
    doi = "10.1103/PhysRevD.93.055004",
    journal = "Phys. Rev. D",
    volume = "93",
    number = "5",
    pages = "055004",
    year = "2016"
}

@article{Chang:1999nh,
    author = "Chang, Sanghyeon and Hisano, Junji and Nakano, Hiroaki and Okada, Nobuchika and Yamaguchi, Masahiro",
    title = "{Bulk standard model in the Randall-Sundrum background}",
    eprint = "hep-ph/9912498",
    archivePrefix = "arXiv",
    reportNumber = "TU-581, KEK-TH-665, NIIG-DP-99-3",
    doi = "10.1103/PhysRevD.62.084025",
    journal = "Phys. Rev. D",
    volume = "62",
    pages = "084025",
    year = "2000"
}

@article{Gherghetta:2000qt,
    author = "Gherghetta, Tony and Pomarol, Alex",
    title = "{Bulk fields and supersymmetry in a slice of AdS}",
    eprint = "hep-ph/0003129",
    archivePrefix = "arXiv",
    reportNumber = "CERN-TH-2000-081, UNIL-IPT-00-06",
    doi = "10.1016/S0550-3213(00)00392-8",
    journal = "Nucl. Phys. B",
    volume = "586",
    pages = "141--162",
    year = "2000"
}

@article{Contino:2003ve,
    author = "Contino, Roberto and Nomura, Yasunori and Pomarol, Alex",
    title = "{Higgs as a Holographic Pseudo Goldstone Boson}",
    eprint = "hep-ph/0306259",
    archivePrefix = "arXiv",
    reportNumber = "FT-UAM-03-11, FERMILAB-PUB-03-195-T, UAB-FT-549",
    doi = "10.1016/j.nuclphysb.2003.08.027",
    journal = "Nucl. Phys. B",
    volume = "671",
    pages = "148--174",
    year = "2003"
}

@article{Schmaltz:2002wx,
    author = "Schmaltz, Martin",
    editor = "Bentvelsen, S. and de Jong, P. and Koch, J. and Laenen, Eric",
    title = "{Physics beyond the standard model (theory): Introducing the little Higgs}",
    eprint = "hep-ph/0210415",
    archivePrefix = "arXiv",
    reportNumber = "BUHEP-02-35, BUPUB-02-35",
    doi = "10.1016/S0920-5632(03)01409-9",
    journal = "Nucl. Phys. B Proc. Suppl.",
    volume = "117",
    pages = "40--49",
    year = "2003"
}

@article{Grimus:2007if,
    author = "Grimus, W. and Lavoura, L. and Ogreid, O. M. and Osland, P.",
    title = "{A Precision constraint on multi-Higgs-doublet models}",
    eprint = "0711.4022",
    archivePrefix = "arXiv",
    primaryClass = "hep-ph",
    reportNumber = "UWTHPH-2007-28",
    doi = "10.1088/0954-3899/35/7/075001",
    journal = "J. Phys. G",
    volume = "35",
    pages = "075001",
    year = "2008"
}

@article{Angelescu:2015uiz,
    author = "Angelescu, Andrei and Djouadi, Abdelhak and Moreau, Gr{\'e}gory",
    title = "{Scenarii for interpretations of the LHC diphoton excess: two Higgs doublets and vector-like quarks and leptons}",
    eprint = "1512.04921",
    archivePrefix = "arXiv",
    primaryClass = "hep-ph",
    reportNumber = "LPT-ORSAY-15-99",
    doi = "10.1016/j.physletb.2016.02.064",
    journal = "Phys. Lett. B",
    volume = "756",
    pages = "126--132",
    year = "2016"
}

@article{Aguilar-Saavedra:2002phh,
    author = "Aguilar-Saavedra, J. A.",
    title = "{Effects of mixing with quark singlets}",
    eprint = "hep-ph/0210112",
    archivePrefix = "arXiv",
    reportNumber = "FISIST-16-2002-CFIF",
    doi = "10.1103/PhysRevD.69.099901",
    journal = "Phys. Rev. D",
    volume = "67",
    pages = "035003",
    year = "2003",
    note = "[Erratum: Phys.Rev.D 69, 099901 (2004)]"
}

@article{Benbrik:2025kvz,
    author = "Benbrik, R. and Berrouj, M. and Boukidi, M. and Ech-chaouy, M. and Kahime, K. and Salime, K.",
    title = "{Relaxing vector-like top quark mass limits through exotic decays in the type-II two-Higgs-doublet model}",
    doi = "10.1140/epjc/s10052-025-15047-3",
    journal = "Eur. Phys. J. C",
    volume = "85",
    number = "11",
    pages = "1275",
    year = "2025"
}

@misc{CMS:2025zwi,
    author = "Hayrapetyan, Aram and others",
    collaboration = "CMS",
    title = "{Search for single production of a vector-like T quark decaying to a top quark and a neutral scalar boson in the lepton+jets final state in proton-proton collisions at $\sqrt{s}$ = 13 TeV}",
    eprint = "2510.25874",
    archivePrefix = "arXiv",
    primaryClass = "hep-ex",
    reportNumber = "CMS-B2G-23-009, CERN-EP-2025-228",
    month = "10",
    year = "2025"
}

@article{delAguila:2000aa,
    author = "del Aguila, F. and Perez-Victoria, M. and Santiago, Jose",
    title = "{Effective description of quark mixing}",
    eprint = "hep-ph/0007160",
    archivePrefix = "arXiv",
    reportNumber = "UG-FT-116-00, MIT-CTP-2996",
    doi = "10.1016/S0370-2693(00)01071-6",
    journal = "Phys. Lett. B",
    volume = "492",
    pages = "98--106",
    year = "2000"
}

@article{CMS:2024bni,
    author = "Hayrapetyan, Aram and others",
    collaboration = "CMS",
    title = "{Review of searches for vector-like quarks, vector-like leptons, and heavy neutral leptons in proton{\textendash}proton collisions at {\ensuremath{\sqrt{}}}s=13 TeV at the CMS experiment}",
    eprint = "2405.17605",
    archivePrefix = "arXiv",
    primaryClass = "hep-ex",
    reportNumber = "CMS-EXO-23-006, CERN-EP-2024-095",
    doi = "10.1016/j.physrep.2024.09.012",
    journal = "Phys. Rept.",
    volume = "1115",
    pages = "570--677",
    year = "2025"
}

@article{Sjostrand:2014zea,
    author = {Sj{\"o}strand, Torbj{\"o}rn and Ask, Stefan and Christiansen, Jesper R. and Corke, Richard and Desai, Nishita and Ilten, Philip and Mrenna, Stephen and Prestel, Stefan and Rasmussen, Christine O. and Skands, Peter Z.},
    title = "{An introduction to PYTHIA 8.2}",
    eprint = "1410.3012",
    archivePrefix = "arXiv",
    primaryClass = "hep-ph",
    reportNumber = "LU-TP-14-36, MCNET-14-22, CERN-PH-TH-2014-190, FERMILAB-PUB-14-316-CD, DESY-14-178, SLAC-PUB-16122",
    doi = "10.1016/j.cpc.2015.01.024",
    journal = "Comput. Phys. Commun.",
    volume = "191",
    pages = "159--177",
    year = "2015"
}

@article{deFavereau:2013fsa,
    author = "de Favereau, J. and Delaere, C. and Demin, P. and Giammanco, A. and Lema{\^\i}tre, V. and Mertens, A. and Selvaggi, M.",
    collaboration = "DELPHES 3",
    title = "{DELPHES 3, A modular framework for fast simulation of a generic collider experiment}",
    eprint = "1307.6346",
    archivePrefix = "arXiv",
    primaryClass = "hep-ex",
    doi = "10.1007/JHEP02(2014)057",
    journal = "JHEP",
    volume = "02",
    pages = "057",
    year = "2014"
}

@article{Cacciari:2008gp,
    author = "Cacciari, Matteo and Salam, Gavin P. and Soyez, Gregory",
    title = "{The anti-$k_t$ jet clustering algorithm}",
    eprint = "0802.1189",
    archivePrefix = "arXiv",
    primaryClass = "hep-ph",
    reportNumber = "LPTHE-07-03",
    doi = "10.1088/1126-6708/2008/04/063",
    journal = "JHEP",
    volume = "04",
    pages = "063",
    year = "2008"
}

@article{Conte:2013mea,
    author = "Conte, Eric and Fuks, Benjamin",
    editor = "Wang, Jianxiong",
    title = "{MadAnalysis 5: status and new developments}",
    eprint = "1309.7831",
    archivePrefix = "arXiv",
    primaryClass = "hep-ph",
    reportNumber = "CERN-PH-TH-2013-236",
    doi = "10.1088/1742-6596/523/1/012032",
    journal = "J. Phys. Conf. Ser.",
    volume = "523",
    pages = "012032",
    year = "2014"
}

@article{Chen:2016btl,
  author        = {Tianqi Chen and Carlos Guestrin},
  title         = {XGBoost: A Scalable Tree Boosting System},
  journal       = {arXiv preprint arXiv:1603.02754},
  year          = {2016},
  archivePrefix = {arXiv},
  eprint        = {1603.02754},
  primaryClass  = {cs.LG}
}

@misc{ross2014introduction,
  title={Introduction to probability and statistics for engineers and scientists},
  author={Ross, Sheldon M},
  year={2014},
  publisher={Academic Press},
  edition={5}
}

@article{Matsedonskyi:2014mna,
    author = "Matsedonskyi, Oleksii and Panico, Giuliano and Wulzer, Andrea",
    title = "{On the Interpretation of Top Partners Searches}",
    eprint = "1409.0100",
    archivePrefix = "arXiv",
    primaryClass = "hep-ph",
    reportNumber = "CERN-PH-TH-2014-160",
    doi = "10.1007/JHEP12(2014)097",
    journal = "JHEP",
    volume = "12",
    pages = "097",
    year = "2014"
}

@article{Roy:2020fqf,
    author = "Roy, Avik and Nikiforou, Nikiforos and Castro, Nuno and Andeen, Timothy",
    title = "{Novel interpretation strategy for searches of singly produced vectorlike quarks at the LHC}",
    eprint = "2003.00640",
    archivePrefix = "arXiv",
    primaryClass = "hep-ph",
    doi = "10.1103/PhysRevD.101.115027",
    journal = "Phys. Rev. D",
    volume = "101",
    number = "11",
    pages = "115027",
    year = "2020"
}

@article{Bevilacqua:2022ozv,
    author = "Bevilacqua, Giuseppe and Lupattelli, Michele and Stremmer, Daniel and Worek, Malgorzata",
    title = "{Study of additional jet activity in top quark pair production and decay at the LHC}",
    eprint = "2212.04722",
    archivePrefix = "arXiv",
    primaryClass = "hep-ph",
    reportNumber = "P3H-22-121, TTK-22-44",
    doi = "10.1103/PhysRevD.107.114027",
    journal = "Phys. Rev. D",
    volume = "107",
    number = "11",
    pages = "114027",
    year = "2023"
}

@article{Bevilacqua:2009zn,
    author = "Bevilacqua, G. and Czakon, M. and Papadopoulos, C. G. and Pittau, R. and Worek, M.",
    title = "{Assault on the NLO Wishlist: pp ---{\ensuremath{>}} t anti-t b anti-b}",
    eprint = "0907.4723",
    archivePrefix = "arXiv",
    primaryClass = "hep-ph",
    reportNumber = "PITHA-09-18, WUB-09-09",
    doi = "10.1088/1126-6708/2009/09/109",
    journal = "JHEP",
    volume = "09",
    pages = "109",
    year = "2009"
}

@article{Bevilacqua:2022nrm,
    author = "Bevilacqua, Giuseppe and Hartanto, Heribertus Bayu and Kraus, Manfred and Nasufi, Jasmina and Worek, Malgorzata",
    title = "{NLO QCD corrections to full off-shell production of $ t\overline{t}Z $ including leptonic decays}",
    eprint = "2203.15688",
    archivePrefix = "arXiv",
    primaryClass = "hep-ph",
    reportNumber = "TTK-22-02, P3H-22-002, CAVENDISH-HEP-22/03",
    doi = "10.1007/JHEP08(2022)060",
    journal = "JHEP",
    volume = "08",
    pages = "060",
    year = "2022"
}

@article{Campbell:2012dh,
    author = "Campbell, John M. and Ellis, R. Keith",
    title = "{$t \bar{t} W^{+-}$ production and decay at NLO}",
    eprint = "1204.5678",
    archivePrefix = "arXiv",
    primaryClass = "hep-ph",
    reportNumber = "FERMILAB-PUB-12-109-T",
    doi = "10.1007/JHEP07(2012)052",
    journal = "JHEP",
    volume = "07",
    pages = "052",
    year = "2012"
}

@article{Stremmer:2021bnk,
    author = "Stremmer, Daniel and Worek, Malgorzata",
    title = "{Production and decay of the Higgs boson in association with top quarks}",
    eprint = "2111.01427",
    archivePrefix = "arXiv",
    primaryClass = "hep-ph",
    reportNumber = "P3H-21-064, TTK-21-38",
    doi = "10.1007/JHEP02(2022)196",
    journal = "JHEP",
    volume = "02",
    pages = "196",
    year = "2022"
}

@misc{Benbrik:2026zjv,
    author = "Benbrik, R. and Berrouj, M. and Boukidi, M. and Chatoui, H. and Ech-chaouy, M. and Kahime, K. and Salime, K.",
    title = "{Single Production of a Vector-Like Top as a Probe of Charged Higgs Bosons at a Muon-Proton Collider}",
    eprint = "2601.07758",
    archivePrefix = "arXiv",
    primaryClass = "hep-ph",
    month = "1",
    year = "2026"
}

@article{BENBRIK2026117436,
title = {Impact of hidden heavy Higgs channels of VLB-Quarks below 1 TeV in 2HDM},
journal = {Nuclear Physics B},
pages = {117436},
year = {2026},
issn = {0550-3213},
doi = {https://doi.org/10.1016/j.nuclphysb.2026.117436},
url = {https://www.sciencedirect.com/science/article/pii/S0550321326001446},
author = {R. Benbrik and M. Berrouj and M. Boukidi and M. Ech-Chaouy and K. Kahime and K. Salime},
}

@article{Yang:2025zbp,
    author = "Yang, Shuo and Wang, Baoxia and Zhu, Pengxuan",
    title = "{Exploring vector-like B-quark pair production at CLIC in fully hadronic final states}",
    eprint = "2510.10237",
    archivePrefix = "arXiv",
    primaryClass = "hep-ph",
    doi = "10.1140/epjc/s10052-026-15400-0",
    journal = "Eur. Phys. J. C",
    volume = "86",
    number = "3",
    pages = "217",
    year = "2026"
}

@article{Backovic:2014uma,
    author = "Backovi{\'c}, Mihailo and Flacke, Thomas and Lee, Seung J. and Perez, Gilad",
    title = "{LHC Top Partner Searches Beyond the 2 TeV Mass Region}",
    eprint = "1409.0409",
    archivePrefix = "arXiv",
    primaryClass = "hep-ph",
    doi = "10.1007/JHEP09(2015)022",
    journal = "JHEP",
    volume = "09",
    pages = "022",
    year = "2015"
}

@article{Atre:2011ae,
    author = "Atre, Anupama and Azuelos, Georges and Carena, Marcela and Han, Tao and Ozcan, Erkcan and Santiago, Jose and Unel, Gokhan",
    title = "{Model-Independent Searches for New Quarks at the LHC}",
    eprint = "1102.1987",
    archivePrefix = "arXiv",
    primaryClass = "hep-ph",
    reportNumber = "FERMILAB-PUB-11-910-T",
    doi = "10.1007/JHEP08(2011)080",
    journal = "JHEP",
    volume = "08",
    pages = "080",
    year = "2011"
}

@article{Nutter:2012an,
    author = "Nutter, Joseph and Schwienhorst, Reinhard and Walker, Devin G. E. and Yu, Jiang-Hao",
    title = "{Single Top Production as a Probe of B-prime Quarks}",
    eprint = "1207.5179",
    archivePrefix = "arXiv",
    primaryClass = "hep-ph",
    reportNumber = "MSUHEP-120723",
    doi = "10.1103/PhysRevD.86.094006",
    journal = "Phys. Rev. D",
    volume = "86",
    pages = "094006",
    year = "2012"
}

@article{Banerjee:2016wls,
	author = "Banerjee, Shankha and Barducci, Daniele and B{\'e}langer, Genevi{\`e}ve and Delaunay, C{\'e}dric",
	title = "{Implications of a High-Mass Diphoton Resonance for Heavy Quark Searches}",
	eprint = "1606.09013",
	archivePrefix = "arXiv",
	primaryClass = "hep-ph",
	reportNumber = "LAPTH-033-16",
	doi = "10.1007/JHEP11(2016)154",
	journal = "JHEP",
	volume = "11",
	pages = "154",
	year = "2016"
}

@article{Liu:2024hvp,
	author = "Liu, Yao-Bei and Hu, Bo and Li, Chao-Zheng",
	title = "{Single production of vectorlike quarks with charge 5/3 at the 14 TeV LHC}",
	eprint = "2402.01248",
	archivePrefix = "arXiv",
	primaryClass = "hep-ph",
	doi = "10.1016/j.nuclphysb.2024.116667",
	journal = "Nucl. Phys. B",
	volume = "1007",
	pages = "116667",
	year = "2024"
}

@article{Zhang:2024nto,
	author = "Zhang, Yan-Ju and Zhu, Yong-Tao and Han, Lin and Bi, Yan-Ping and Liu, Tai-Gang",
	title = "{Searching for the vector-like X-quark at the future electron{\textendash}proton colliders}",
	doi = "10.1140/epjc/s10052-024-13475-1",
	journal = "Eur. Phys. J. C",
	volume = "84",
	number = "11",
	pages = "1184",
	year = "2024"
}
	
\end{document}